\documentclass{emulateapj}
\usepackage{amssymb}
\shorttitle{Dust and Gas in W3-SE}
\shortauthors{Zhu et al.}
\begin{document}
\title{Dust and HCO$^+$ Gas in the Star Forming Core W3-SE}
\author{Lei Zhu\altaffilmark{1,3}, M. C. H. Wright\altaffilmark{2},
Jun-Hui Zhao\altaffilmark{1} and Yuefang Wu\altaffilmark{3}}
\altaffiltext{1}{Harvard-Smithsonian Center for Astrophysics, 60
Garden Street, Cambridge, MA
 02138; lzhu@cfa.harvard.edu}
\altaffiltext{2}{Department of Astronomy, University of California,
Berkeley, Berkeley, CA 94720} \altaffiltext{3}{Department of
Astronomy, Peking University, Beijing, China}

\begin{abstract}

We report new results from recent Combined Array for Research in
Millimeter-wave Astronomy (CARMA) observations of both continuum and
HCO$^+$(1-0) line emission at $\lambda$~3.4 mm from W3-SE, a molecular
core of intermediate mass, together with the observations of continuum
emission at $\lambda$~1.1 and 0.85/0.45 mm with the Submillimeter Array
(SMA) and the James Clerk Maxwell Telescope (JCMT), respectively. A
continuum emission core elongated from SE to NW, with a size of
$\sim$10$\arcsec$, has been observed at the millimeter and
submillimeter wavelengths. The dust core has been resolved into a
double source with the SMA at $\lambda$~1.1 mm. The angular separation
between the two components is $\sim$4$\arcsec$. Together with the
measurements from the {\it Spitzer Space Telescope} and the {\it
Midcourse Space Experiment} (MSX) at mid-IR wavelengths, we determined
the spectral energy distribution (SED) of the continuum emission from
W3-SE and fit it with a thermal dust emission model. Our best fitting
of the SED suggests the presence of two dust components with different
temperatures. The emission at mm/submm wavelengths is dominated by a
major component which is characterized by a temperature of $T_{\rm
d}$=41$\pm$6 K with a mass of 65$\pm$10 M$_\sun$. In addition, there is
a weaker hot component ($\sim$ 400 K) which accounts for emission in
the mid-IR, suggesting that a small fraction of dust has been heated by
newly formed stars. We also imaged the molecular core in the
HCO$^+$(1-0) line using CARMA at an angular resolution
$\sim$6$\arcsec$. In the central region of $\sim$50$\arcsec$, the
integrated HCO$^+$(1-0) line emission shows a main component A that
coincides with the dust core, as well as two sub-structures B and C
which are located N and SE of the dust core, respectively. With the
CARMA observations, we have verified the presence of a blue-dominated
double peak profile toward this core. The line profile cannot be
explained by infall alone. The broad velocity wings of the line profile
suggest that other kinematics such as outflows within the central
6$\arcsec$ of the core likely dominate the resulting spectrum. The
kinematics of the sub-structures of B and C suggest that the molecular
gas outside the main component A appears to be dominated by the bipolar
outflow originated from the dust core with a dynamical age of
$>$3$\times$10$^4$ yr. Our analysis, based on the observations at
wavelengths from millimeter, submillimeter to mid-IR suggests that the
molecular core W3-SE hosts a group of newly formed young stars and
protostars.
\end{abstract}


\section{Introduction}

Among the dynamical processes in star formation regions, infall appears
to be critical in understanding the processes of mass accumulation in
star formation cores and the growth of protostars. The protostellar
infall rate depends on the surface density of the protostellar core,
which usually can be determined from observations of self-absorptions
of the embedded molecular clumps in it. However, it is difficult to
unambiguously identify infall based on low angular-resolution
observations because the spectral features of a complex with bipolar
molecular outflows, rotating structures or multiple cores can produce
similar characteristics to those expected from self-absorption in
infalling clumps. In many young star formation regions that show
spectral signatures of infall from single-dish observations, the
single-dish spectral features often turn out to be dominated by
outflows and other kinematic features with further high-resolution
observations using interferometer arrays.

W3-SE is a dense molecular core located 2 kpc away \citep{blit82}, SE
of the well-known massive star formation region W3-Main. The W3-SE core
was first imaged by \cite{tief98} in the NH$_3$(1,1) and (2,2) lines.
The low kinetic temperature of 25 K based on the NH$_3$ line ratio
\citep{tief98} and the absence of IRAS emission \citep{jiji99} implied
that this core is probably in an early evolutionary stage. A virial
mass of $\sim$300 M$_\sun$ was inferred from the line widths,
indicating that this core may be a birthplace for intermediate- or
high-mass stars. However, to date there has been no evidence for OB
stars present in the core because of the non-detection from the radio
continuum emission at 6 cm with the VLA \citep{tief98}. On the other
hand, a possible infall signature in W3-SE was suggested by \cite{wu07}
based on observations of multiple molecular lines with the IRAM 30-m
telescope. Both the HCO$^+$ (1-0) and (3-2) lines show an asymmetric
double profile characterized by a primary peak blue-shifted with
respect to a peak velocity of the optically thin lines such as
C$^{18}$O(1-0), while a secondary peak is red-shifted. The possible
signature of infall in W3-SE suggested from the single-dish
observations needs further verification with improved, high resolution
observations. Besides, \cite{ruch07} reported a detection of a mid-IR
point source in the region of W3-SE and determined the flux densities
in the four bands of the Infrared Array Camera (IRAC) on the {\it
Spitzer Space Telescope}. The color-color diagram in their paper
indicates that this source is most likely to be a YSO (Class I or
earlier) with large excess emissions at longer wavelengths.

In this paper, we present new results based on observations at
$\lambda$~3.4, 1.1, 0.85/0.45 mm with the Combined Array for Research
in Millimeter-wave Astronomy (CARMA), the Submillimeter Array
(SMA)\footnote{The Submillimeter Array is a joint project between the
Smithsonian Astrophysical Observatory and the Academia Sinica Institute
of Astronomy and Astrophysics and is funded by the Smithsonian
Institution and the Academia Sinica.} and the James Clerk Maxwell
Telescope (JCMT)\footnote{The James Clerk Maxwell Telescope is operated
by the Joint Astronomy Centre on behalf of the Particle Physics and
Astronomy Research Council of the United Kingdom, the Netherlands
Organisation for Scientific Research, and the National Research Council
of Canada.}, respectively, along with the data at mid-IR wavelengths
obtained from both the {\it Spitzer Space Telescope} and {\it Midcourse
Space Experiment} (MSX) archives. \S2 describes observations and data
reductions, \S3 presents results from the analysis of the continuum
emission at multiple wavelengths and the CARMA observations of the
HCO$^+$(1-0) line, and \S4 discusses the results and their implications
for astrophysics. Finally, \S5 gives the summary and conclusions.

\section{Observations and Data Reduction}

\subsection{CARMA Observations at $\lambda$~3.4 mm}

W3-SE was observed at $\lambda$~3.4 mm with the CARMA telescope in the
D-configuration on June 24th and June 27th, 2008, with an angular
resolution of $\sim6\arcsec$. The observations were centered at
RA(J2000) = 02$^{\rm h}$25$^{\rm m}$54$^{\rm s}$.50, DEC(J2000) =
62$\arcdeg$04$\arcmin$11$\arcsec$.0, which is used as the reference
position in the following analysis. Two broad bands with BW of 500 MHz
were set at 86 and 89 GHz to image the continuum emission. Two USB
narrow bands with BW of 8 and 32 MHz were tuned to the HCO$^+$(1-0)
line at 89.188 GHz, and the two LSB narrow bands were used for the
SiO(2-1) line at 86.243 GHz. The highest velocity resolution is 0.42 km
s$^{-1}$ for the 8-MHz narrow-band data with 63 spectral channels. The
observations were Doppler-tracked at $V_{\rm LSR}$ = --38.9 km
s$^{-1}$.

The observations include two tracks. The first one lasted for 5 hours
using 14 antennas with a 19-point mosaic, and the second one was taken
for 4 hours using 13 antennas with a 7-point mosaic. Because the first
track was on a maintenance day, several antennas were not used in parts
of the track. The corrupted uv data were flagged during the off-line
data reduction. The second track was carried out in good weather, and
the data quality is excellent with $T_{\rm sys}$ between 150 and 200 K.

All the data reductions and imaging processes were carried out in {\it
MIRIAD} \citep{saul95}. The data from the two tracks were calibrated
separately with the calibrators listed in Table 1. The bandpass and
complex gain calibrations were processed following the standard
procedure for the CARMA data reduction. Flux density scales were
calibrated using Uranus (7.2 Jy) and MWC 349 (1.3 Jy) at $\lambda$~3.4
mm for the two tracks, respectively. After calibration the two datasets
were merged and images were made with a linear mosaic algorithm using
the program {\it INVERT}.

\subsection{Combining CARMA Data with IRAM 30-m Data}

The shortest baseline of the CARMA observations is 2.7 k$\lambda$. To
eliminate the effects of missing short spacing, we combined the CARMA
and IRAM 30-m telescope data which cover the central
45$\arcsec$$\times$45$\arcsec$ region \citep{wu07}. We used the {\it
MIRIAD} program {\it IMMERGE} to linearly combine the single-dish image
and the cleaned CARMA image. The uv ranges sampled by the two
observations are well overlapped, which are used to determine the
scaling factor for the flux density of the single-dish data. The
amplitude of the IRAM 30-m data was re-scaled by applying the scaling
factor determined by comparing the amplitude values of the overlapped
visibilities in the FFT domain. There is a boundary effect on the edge
of the combined image of 45$\arcsec$$\times$45$\arcsec$ due to limited
spatial coverage of the IRAM observations. The effect is negligible in
our analysis of the line emission in the central
20$\arcsec$$\times$20$\arcsec$ region.

\subsection{SMA Observations at $\lambda$~1.1 mm}

We also made observations at $\lambda$~1.1 mm with the SMA in the
compact array with an angular resolution of $\sim$2.5$\arcsec$. The SMA
observations were centered at the same position as that used in the
CARMA observations. The frequencies were tuned to 266.7/276.7 GHz for
LSB/USB with bandwidths of 2 GHz for each of the side bands. The
typical T$_{\rm sys}$ in the observations is 180-200 K.

The data reduction and imaging were made using {\it MIRIAD}. The
bandpass shapes and complex gains were calibrated using QSOs 3C454.3
and 0244+624, respectively. The flux density scale was bootstrapped
from Uranus using the SMA planet model. The continuum image of W3-SE
was constructed by combining the visibility data from the two side
bands with a rms noise of 1.5 mJy.

\subsection{JCMT Observations at $\lambda$~0.85/0.45 mm}

The W3-SE region was observed using JCMT with SCUBA at $\lambda$~0.85
and 0.45 mm on September 15, 2004 (PID M04BC17). The data were
calibrated and combined into intensity images with the ORAC-DR
pipeline. The FWHM beam sizes of the JCMT images are 14.5$\arcsec$ and
7.5$\arcsec$ at $\lambda$~0.85 and 0.45 mm, respectively. The JCMT
images were re-grided with the program {\it REGRID} in {\it MIRIAD} to
align the coordinate axes with those in the CARMA images.

\subsection{{\it Spitzer} and MSX Data at Mid-IR}

The entire region of W3 was observed by R. Gehrz (the PI in GTO program
PID 127) with the IRAC \citep{fazi04} on January 10, 2004 and the
Multiband Imaging Photometer for {\it Spitzer} \citep[MIPS;][]{riek04}
on January 10, 2004 on board the {\it Spitzer Space telescope}. The
results have been published by \cite{ruch07}, and a point source which
is detected in all four IRAC bands in the W3-SE region was reported. In
order to search for additional weaker mid-IR sources to identify the
possible counterparts of the radio sources that have been detected at
mm/submm wavelengths, we have further investigated the W3-SE region by
re-reducing the archive data of IRAC and MIPS (GTO program PID 127).

The basic calibrated data (BCD) products of IRAC are provided by the
{\it Spitzer} Science Center (SCC) with the IRAC pipeline version
S18.7.0. A post-BCD script {\it hdr\_mask.pl} provided by SSC was used
to fix the potential saturation problem. The background correction and
mosaic were made with the version 18.2.0 of the post-BCD reduction
software MOPEX \citep{mako05}, producing a mosaic image with
1.22$\arcsec$ per pixel. The point-source extraction and the aperture
photometry were done with the package APEX in MOPEX. Point-spread
function (PSF) fittings were carried out in the process of detecting
point-like sources, and those with signal-to-noise ratio higher than 6
are identified as point sources. An aperture with radius of
2.4$\arcsec$ and an annulus between 2.4$\arcsec$ and 7.3$\arcsec$ were
used for the aperture photometry to determine the flux densities of the
point sources. Aperture correction, color correction and pixel-phase
correction were carried out according to the procedures described in
the IRAC Data Handbook (version 3.0). In addition, the flux density of
the W3-SE core was determined using the procedures of aperture
photometry for extended sources with a radius of 8.5$\arcsec$ in order
to compare with the flux densities of millimeter/submm in the same
region and to determine the SED. We noted that it was difficult to make
a reliable annulus to represent the background levels due to the
background/foreground stars. Therefore, to avoid obvious stellar
emission, we measured flux densities in several regions which are close
to W3-SE to calculate the background levels. An overall fractional
error of $\sim$15\% was assessed for the photometry, dominated by the
uncertainties in the corrections for the background/foreground emission
levels ($\sim$10\%) and the aperture correction ($\sim$10\%).

The BCD products of MIPS were pre-processed by the MIPS pipeline
version S16.1.0. The W3-SE region is badly saturated in the 70 and 160
$\micron$ images due to the strong emission of W3-Main; thus only the
24-$\micron$ data were used. A post-BCD script {\it flatfield.pl}
provided by SSC was used to make an extra flat-fielding to remove the
potential residual background gradient, and then the background
correction and mosaic were made with MOPEX. The aperture photometry was
carried out with the program {\it aperture} in MOPEX with an aperture
radius of 13.5$\arcsec$, and color/aperture corrections were made
following the MIPS Data Handbook (version 3.3.1). An uncertainty of
15\% in the flux densities is dominated by the errors in corrections
for the background/foreground emission ($\sim$10\%) and the
color/aperture corrections.

In addition, we also investigated the MSX images of W3-SE in all the
four bands (8.3, 16.7, 12.1 and 21.3 $\micron$) with an angular
resolution of $\sim$20$\arcsec$. We re-grided all of the {\it Spitzer}
and MSX images in {\it MIRIAD} to align the coordinate axes with those
in the CARMA images. The photometry was carried out with the similar
methods as those for the IRAC images. The uncertainties of the
photometry on the MSX images are large ($\sim$40\%) due to poor
sensitivities, low angular resolutions and severe contamination from
background/foreground emission.

\section{Results}
\subsection{Continuum Emission}
\subsubsection{CARMA $\lambda$~3.4 mm}

Figure 1(a) shows the CARMA image of W3-SE at $\lambda$~3.4 mm,
revealing an un-resolved dust core plus a tail-like structure extended
in SE. We determined a peak position ($\delta$(R.A.) =
--5.5$\pm$0.2$\arcsec$ and $\delta$(DEC) = --0.9$\pm$0.2$\arcsec$) and
peak intensity (24$\pm$2 mJy beam$^{-1}$) for the continuum emission
from the dust core based on a Gaussian fitting to the data. Integrated
over the core region of 15$\arcsec$$\times$15$\arcsec$, a total flux
density of 42$\pm$10 mJy was determined.

\subsubsection{SMA $\lambda$~1.1 mm}

Figure 1(b) shows the SMA image at $\lambda$~1.1 mm. The dust core has
been resolved into two continuum emission components (SMA-1 and -2),
and the overall morphology of the W3-SE dust core shows an elongation
from NW to SE. The primary (SMA-1) and secondary (SMA-2) components of
the double source are located NW and SE of the CARMA peak position.
From fits with two Gaussian components and a linear background, we
obtained peak intensities of 0.21$\pm$0.01 and 0.17$\pm$0.01 Jy
beam$^{-1}$ for SMA-1 and SMA-2, respectively. The angular separation
of the double components is $\sim$4$\arcsec$. The total flux densities
and sizes of the double components are listed in Table.2. The peak
position of the emission at $\lambda$~3.4 mm observed with CARMA agrees
well with the centroid position of the double components observed with
the SMA. We integrated the continuum flux density from the core in the
SMA 1.1-mm image using {\it IMAGR} in {\it AIPS} by adding a zero
spacing flux density of 2 Jy which was derived from multiple Gaussian
fits to the visibility data, and determined that the total continuum
flux density at 1.1 mm from the W3-SE core
($\sim$15$\arcsec$$\times$15$\arcsec$) is 1.4$\pm$0.4 Jy. The large
uncertainty in the flux density includes all possible error sources,
dominated by the missing short spacings in the SMA observations.

\subsubsection{JCMT $\lambda$~0.45/0.85 mm}

Figures 1(c) and 1(d) show the continuum images observed with JCMT at
$\lambda$~0.85 and 0.45 mm, respectively. With the angular resolutions
14.5$\arcsec$ and 7.5$\arcsec$ at $\lambda$~0.85 and 0.45 mm, the
emission at submm wavelengths from W3-SE has been slightly resolved,
showing an extended, $\sim$ 1$\arcmin$ halo, which is better seen in
the JCMT image at $\lambda$~0.85 mm (Figure 1(c)). An elongated core,
$\sim$ 15$\arcsec$, is clearly shown at $\lambda$~0.45 mm in Fig.1d.
The core-halo morphology of the continuum emission observed at submm
with the JCMT is consistent with CARMA and SMA observations at
$\lambda$~3.4 and 1.1 mm, suggesting that the double compact continuum
core (SMA-1 and SMA-2) revealed by the SMA is encompassed by a halo.
Gaussian fits to the JCMT continuum images give peak positions
(--8.2$\arcsec$$\pm$0.4$\arcsec$, 1.2$\arcsec$$\pm$0.6$\arcsec$) and
(--8.8$\arcsec$$\pm$0.4$\arcsec$, 0.5$\arcsec$$\pm$0.4$\arcsec$) for
the $\lambda$~0.45 and 0.85 mm emission, respectively. Within the JCMT
beams, these two positions agree with the position of the continuum
peak observed with CARMA at $\lambda$~3.4 mm and the centroid position
of the double components derived from the SMA observations at
$\lambda$~1.1 mm.

From the JCMT images, integrating over emission region with signals
greater than 5 $\sigma$, we derived a total flux density 8$\pm$1 and
75$\pm$8 Jy at $\lambda$~0.85 and 0.45 mm, respectively. All the
measurements of flux densities at millimeter and submm wavelengths are
summarized in Table 2.

\subsubsection{MIPS and IRAC at Mid-Infrared}

In the MIPS image at 24 $\micron$ (Figure 2(a)), the direction of the
elongation in the 24-$\micron$ emission is in NW-SE
(P.A.$\approx$--25$\arcdeg$). In comparison with the distribution of
the emission at millimeters/submm, we noticed that in the region
located E of the SMA double sources there is a large excess emission at
24 $\micron$, indicating a displacement between the hot and cold dust
emission in the W3-SE core.

Based on the IRAC images, a group of discrete emission sources were
revealed in the core of W3-SE (see the 3.6 $\micron$ image in Figure
2(b)). By fitting PSF, three point sources were detected in the central
core region of W3-SE. One of them has been reported in \cite{ruch07} as
the IRAC point source 56 (hereafter IRAC 56), the only source that can
be fitted as a point source in all of the four IRAC bands. The flux
densities of IRAC 56 determined from this paper are in agreement with
the results of \cite{ruch07}. Located $\sim$4$\arcsec$ NW to IRAC 56, a
weak point source (hereafter IRAC 56a) was detected at 4.5, 5.8 and 8.0
$\micron$, corresponding to a weak emission feature at 3.6 $\micron$
which failed to be fitted as a point source. The detection of this
source is also in good agreement with the study of Ruch et al. (Ruch
2009, private communication). This source appears to coincide with the
core SMA-1, showing a large excesses in the emission at the longer
wavelengths (5.8 and 8.0 $\micron$) as compared with that at 4.5 and
3.6 $\micron$. The large excesses in emission at longer wavelengths
suggest that probably a protostar is embedded in SMA-1, and extinctions
at shorter wavelengths become considerable. Located $\sim$2$\arcsec$ SW
to IRAC 56 there appears to be an additional point source (hereafter
IRAC 56b) detected at 4.5 $\micron$, but we failed to fit it as a point
source due to the contamination from the profile wing of the strong
point source IRAC 56 and possible extended PAH or H$_2$ emission. The
nature of this source remains uncertain. In addition, we note that
there appears no significant emission at the mid-IR wavelengths toward
the cold dust core SMA-2.

Following the procedures of the aperture photometry described early, we
determined the mid-IR flux densities from the core region of W3-SE. The
total flux densities of W3-SE at mid-IR wavelengths from the IRAC and
MIPS images, as well as the values from the MSX images, are summarized
in Table 3.

\subsubsection{Dust Content and SED}

Figure 3 shows the spectral energy distribution (SED) of W3-SE covering
the wavelengths from millimeter to mid-IR. We fit the SED with a
thermal dust model to determine the physical parameters of the dust
components in the core of W3-SE. Given the dust temperature $T_{\rm d}$
and the total mass of gas and dust $M_{\rm total}$, the continuum
emission at the frequency $\nu$ from thermal dust in a molecular core
subtending a solid angle $\Omega$ can be described as \citep{wils09}
\begin{equation}
\label{} S_{\nu} = B_{\nu}(T_{\rm
d})(1-e^{-\tau_{\nu}})\Omega
\end{equation}
where
\begin{displaymath}
\tau_{\nu} = M_{\rm
total}\kappa_{\nu0}(\nu/\nu_0)^\beta/gD^2\Omega
\end{displaymath}
In the optically thin case $\tau_{\nu}< 1$, Equation (1) can be
approximately expressed as
\begin{equation}
\label{} S_{\nu} \approx \frac{M_{\rm total}\kappa_{\nu}B_{\nu}(T_{\rm
d})}{gD^2}
\end{equation}
where $\kappa_{\nu}$
is the dust opacity per unit dust mass, $g$ is the density ratio of
H$_2$ gas to dust, $D$ is the distance, and $B_{\nu}(T_{\rm d})$ is the
Planck function with a dust temperature of $T_{\rm d}$. Adopting
$\kappa$(300 GHz) = 1.4 cm$^2$ g$^{-1}$ as the reference value
\citep{osse94}, we assumed a power law of $\kappa_{\nu}$ =
$\kappa_0$($\nu$/$\nu_0$)$^\beta$ \citep{hild83}, where $\beta$ is the
power-law index
for dust opacity. In addition, we adopted $g$ = 100 and $D$ = 2 kpc in
the following analysis. Thus, for each dust component, the thermal dust
model has three free parameters ($M_{\rm total}$, $T_{\rm d}$ and
$\beta$).

Figure 3(a) shows least square (LSQ) fitting to the SED with a total of
thirteen measurements from submm to mid-IR on the assumption of
optically thin dust emission. We found that a model with two thermal
dust components appears to adequately fit the data. The six free
physical parameters of the model can be determined by the LSQ fitting
to the data. Our best fitting suggests that a cold, massive dust
component accounts for the continuum emission at the millimeters/submm
as well as a large fraction of the continuum emission at 24 $\micron$.
A total mass of $M_{\rm total}$ = 70$\pm$7 M$_\sun$, a dust temperature
of $T_{\rm d}$ = 35$\pm$1 K, and a dust opacity index of $\beta
=2.0\pm0.1$ were derived from the fitting for this cold, massive dust
component. In addition, the emission bump at mid-IR requires a
secondary dust component with a small mass but a high temperature. Our
fitting gives $T_{\rm d} = 400^{+70}_{-50}$ K, $\beta$ =
0.5$^{+0.9}_{-0.5}$ and $M_{\rm total} < 5\times10^{-4}$ M$_\sun$ for
the secondary component. We note that the uncertainties of the derived
parameters for the secondary component are large due to the large
uncertainties in the mid-IR measurements, especially the ones from MSX.

Although the optically thin approach is valid for the emission from the
cold dust component at millimeter and submm wavelengths and the
emission from the hot component at mid-IR, the optical depth of the
cold component may become significantly large at the wavelengths of
mid-IR. On the assumption that the steep power law of the dust opacity
($\beta=2$) holds up to the mid-IR, the mean optical depth of
$\tau_\nu\sim$160 at $\lambda$~24 $\micron$ for the cold dust in the
region of the double source is inferred using the values of $T_{\rm
d}$, and $M_{\rm total}$ derived from the optically thin approach.
Thus, the  optical-depth effect can not be ignored for the dust
emission at the shorter wavelengths. Figure 3(b) shows a good fit to
the observed SED data, considering self-absorption in the cold dust
with Equation (1) and assuming that cold component does not overlap the
hot dust emission region. Consequently, in the self-absorption model
the turnover frequency $\nu_{\tau=1}$, where $\tau(\nu)$ = 1, is
shifted to the higher frequency of 1070 GHz as compared with the
value ($\nu_{\tau=1}$ = 990 GHz) derived in the optically thin approach.
The self-absorption model gives a 15\% decrease in $M_{\rm
total}$ and a 25\% increase in $T_{\rm d}$, i.e., $T_{\rm d}$ = 47 K,
and $M_{\rm total}$ = 60 M$_\sun$, if the power-law index $\beta=2$
remains unchanged at mid-IR. In the case that the power-law of the dust
opacity becomes flat, e.g., $\beta$ = 0.5 at the frequencies above 1000
GHz, then $\tau_\nu\sim$3 at $\lambda$~24 $\micron$ and the optical
depth effect at mid-IR becomes not so critical. In any case, the
uncertainties in the determination of $T_{\rm d}$ and $M_{\rm total}$
appears to be mainly due to the uncertainty in the dependence of the
opacity on frequency at mid-IR. Thus, we take the mean values of the
physical parameters derived from the two extreme cases, namely $T_{\rm
d}$ = 41$\pm$6 K, and $M_{\rm total}$ = 65$\pm$10 M$_\sun$. Our result
in dust mass is in good agreement with that of \cite{zinc09}, which
suggests a total mass of 74 M$_\sun$ derived from dust continuum for
W3-SE and a virial mass of 90 M$_\sun$.

On the other hand, we can estimate the extinction from the cold dust
component at mid-IR wavelengths according to recently studies of the
mid-IR extinction law. Based on the high-resolution observations with
the SMA at $\lambda$~1.1 mm, most of the cold dust (over $\sim$85\%) is
concentrated in the two compact mm/submm sources (SMA-1 and -2), the
two possible protostellar cores. Assuming an even distribution of the
dust over the sizes of the two cores (see Column 5 of Table 2), the
average column density of the H$_2$ in the cores would be
1.5$\times$10$^{24}$ cm$^{-2}$. On the basis of the equation
[N(HI)+2N(H$_2$)]/$A_K$ = 1.67$\times$10$^{22}$ cm$^{-2}$ mag$^{-1}$
from \cite{lada09}, such a column density is equivalent to an
extinction $A_K$ = $\sim$200 mag at K band. Assuming a value of 0.5 for
both $A_{IRAC}$/$A_{Ks}$ and $A_{24}$/$A_{Ks}$ based on recent studies
on the mid-IR extinction law \citep[e.g.][]{flah07}, the corresponding
extinctions at the IRAC bands and the $\lambda$~24 $\micron$ would be
$\sim$100 mag. Obviously the light from the sources within the two cold
dust cores are severely obscured at mid-IR wavelengths. The residual of
the cold dust emission ($\sim$15\%, or 0.2 Jy at $\lambda$~1.1 mm)
appears to be distributed outside the two compact cores, corresponding
to a mass of $\sim$10 M$_\sun$. On the assumption of an even
distribution of this extended cold dust component over a region of 0.15
pc$\times$0.15 pc (15$\arcsec$$\times$15$\arcsec$), the column density
N(H$_2$) is about 2.5$\times$10$^{22}$ cm$^{-2}$, corresponding to an
extinction of $\sim$3 mag at K band, or $\sim$1.5 mag at the IRAC bands
and $\lambda$~24 $\micron$. Considering the observed filamentary
structure of the cold dust in the SMA and CARMA observation, the
extinction of the extended cold dust distribution to the mid-IR
emission from W3-SE varies around 1 mag. This extinction may have some
effects on obscuring the emission from the hot dust component depending
on the relative distribution of the extended cold dust with respected
to the hot dust component.

Nevertheless, our fitting to the observed SED of W3-SE suggested the
existence of a dominant cold dust component with a dust temperature of
41$\pm$6 K and a total mass of 65$\pm$10 M$_\sun$. A secondary dust
component which has a little mass but high dust temperature is
necessary to account for the observed emission at IRAC wavelengths,
which is likely heated by the nearby stellar objects (e.g. IRAC 56).
The suggested heating process appears to be also consistent with the
structure of the mid-IR emission characterized as several discrete
compact components as discussed in the previous section (also see
Figure 2).

\subsection{Molecular Lines}
\subsubsection{Non-detection of SiO (2-1) Line}

On the basis of CARMA observations of SiO(2-1) line at 86.243 GHz, we
did not detect SiO
emission at a 3$\sigma$ level ($\sim$0.3 Jy beam$^{-1}$) with a channel
width of 0.42 km s$^{-1}$.  The SiO(2-1) line is not a good tracer for
the outflows in W3-SE.

\subsubsection{Distribution of the HCO$^+$(1-0) Gas}

Figure 4 shows the channel maps of HCO$^+$(1-0) from W3-SE with the
CARMA observations, revealing complex structures in kinematics and
spatial distribution. Strong line emissions are present in the velocity
range from --43.9 to --33.2 km s$^{-1}$ (above 3 $\sigma$), with a gap
between --38.6 and --37.3 km s$^{-1}$. The emission blue-shifted with
respect to the gap is stronger and more extended than the red-shifted
gas. Figure 5(a) shows the HCO$^+$(1-0) line intensity image integrated
over all the line channels, delineating an arc-like morphology in the
distribution of molecular gas. Hereafter, we denote the main component
with the maximum line intensity as A, the northern and SE lobes as B
and C, respectively. The main component A is located at the position
(--5.4$\arcsec$, --1.5$\arcsec$) offset from the reference center,
which coincides with the continuum peak position at $\lambda$~3.4 mm
and the centroid position of the SMA double source. The spectra derived
from the three positions (A, B, and C) are shown in Figure 5(b). The
spectrum at A clearly shows a double-peak line profile; one is at
--40.0 km s$^{-1}$ (or the ``blue peak" hereafter) and the other at
--36.0 km s$^{-1}$ (or ``red peak" hereafter). A spectral dip to
$\sim$0 Jy between the the blue and red peaks is shown, which is
observed as an empty gap between --37.3 and --38.6 km s$^{-1}$ in the
channel maps (Figure 4). In the velocity range from --34.9 to --36.9 km
s$^{-1}$, the red-shifted emission peak appears to coincide near the
position A with the blue-shifted emission peak in the velocity range
from --39.4 to --40.6 km s$^{-1}$, suggesting that the red- and
blue-shifted gas components originate from a common core. At the
position A, the line peak intensities are 4.3 and 2.1 Jy beam$^{-1}$
for the red- and blue-shifted gas in W3-SE, respectively.

The position B is located at (--5.1$\arcsec$, 9.5$\arcsec$), N of the
main core A, appeared to be a sub-core. At the position B the line
emission is much weaker than that at the position A, but the line
intensity ratio of the blue and red peak is $\sim$2, similar to that
observed from the position A. Towards the position C (--1.1$\arcsec$,
--6.6$\arcsec$), the spectrum is dominated by the blue-shifted line
component, while the red-shifted line emission appears to be very weak.

Besides the three relatively compact structures, there appears to be an
extended halo, $\sim$40$\arcsec$ in W3-SE. The total HCO$^+$(1-0) flux
by integrating the line emission above 3$\sigma$ from the entire W3-SE
region is 124 Jy km s$^{-1}$, which underestimates the true flux
because of missing short spacing. The shortest projected baseline of
the CARMA data is 2.7 k$\lambda$, suggesting missing information on the
source structure with a size larger than $\sim$76$\arcsec$. We combined
the CARMA data with the IRAM 30-m data to assess the effects due to
missing short spacing (\S2.2).

Figure 6(a) shows the image constructed from combined CARMA and IRAM
30-m data with the same beam size (6.3$\arcsec$$\times$5.7$\arcsec$,
P.A = --81$\arcdeg$) as that of the CARMA image(Figure 5(a)). The
combined image shows better the emission from the extended halo but
trades off the detailed structures. The total line flux of 260 Jy km
s$^{-1}$ derived from the combined image is about twice as large as
that derived from the CARMA data alone. Towards the peak position A,
the spectrum shows that the intensity peak from the combined data is
$\sim$25\% higher than that of the CARMA data while the intensity at
the spectral dip increases to 0.5 Jy (5 $\sigma$) in the combined image
from $\sim$0 Jy of the CARMA image.

\subsubsection{Outflows}

The broad wings of the line profiles in the HCO$^+$(1-0) spectra
suggest that outflow(s) is active in this molecular cloud. We imaged
the high-velocity portions of the outflow(s) by integrating the blue
spectral wings from --41.4 to --46.7 km s$^{-1}$ and the red spectral
wings from --29.6 to --36.5 km s$^{-1}$. Figure 7(a) shows the blue-
and red-shifted gaseous components in blue and red contours,
respectively. It appears that the red-shifted outflow lobe is located
NW of the continuum core while the blue-shifted outflow lobe shows two
components located SE (C) and N (B) of the continuum core,
respectively. The complex outflow structure implies that there might be
more than one outflow in W3-SE, and/or the direction of the outflow is
close to the line of sight. Cutting along a direction from NW to SE
(marked as the straight line with P.A. = 120$\arcdeg$ in figure.7(a)),
we constructed a position-velocity (P-V) diagram, showing a kinematic
structure which appears to be consistent with a decelerated outflow
\citep{qin08}. The P-V diagram shows only the high velocity components
that originate from the major core A. However, the major axis of the
large-scale outflow is poorly determined due to the poor collimation of
red- and blue-shifted emission.

The line emission feature B mimics an independent molecular core based
on the overlapping blue- and red-shifted outflows shown in Figure 7(a)
although we do not have further evidence either supporting or ruling
out the possibility of it being a low-mass star forming core. The
feature B can also be produced from possible multiple outflows
originated from the major core A. The line emission feature C appears
to be dominated by a blue-shifted outflow lobe originated from the
major core A. To clarify the multiplicity of the outflows in W3-SE,
higher-resolution observations are necessary.

\subsubsection{Possible Infall Signature}

Blue-shifted line profiles were discovered by \cite{wu07} in
HCO$^+$(1-0) and other optically thick lines, which serve as an infall
signature \citep{mard97}. Figure 5(b) shows the spectra at three
positions of the components A, B and C. All the spectral profiles
towards the three regions appear to be characterized by an asymmetric
line profile with a dominated blue-shifted peak and a weaker
red-shifted peak. However, \cite{wu07} also noticed the existence of
the significant line wings in HCO$^+$ lines from their IRAM 30-m
observation. From our CARMA observations, the observed asymmetric
double-peak profiles in spectra towards B and C appear to be due to
outflow. We have shown good evidence for star formation activities
occurring in W3-SE and kinematics of the molecular gas in the central
50$\arcsec$ region which is likely dominated by outflows. We made an
attempt to investigate the possible contribution from infall in the
line profile of the spectrum towards A, the dust core, with a model
including both infall and outflow.

In order to quantitatively differentiate the contributions from
different dynamical processes in the resulting spectral profile, we
utilize the infall model of \cite{myer96} to fit the spectral line
profile. Four parameters: peak optical depths ($\tau_0$), kinetic
temperature ($T_{\rm k}$), velocity dispersion ($\sigma$), and the
characteristic velocity ($V_{\rm in}$) with respect to the system
velocity of ($V_{\rm sys}$), are used in the model to characterize two
layers of infall gas. We found that, indeed, it is difficult to fit the
observed spectral profile using a model with only infall components
because of the broad blue- and red-shifted wings producing sizable
residuals in the fitting.

Thus, we introduced an additional two-layer outflow component with four
more characteristic parameters, namely, brightness temperature $T_{\rm
b,out}$, optical depth ($\tau_{\rm 0,out}$), characteristic velocity
($V_{\rm out}$) with respect to the system, and the velocity dispersion
($\sigma_{\rm out}$) in the Gaussian profile. Similar approach has been
considered in modeling the infall motions in NGC 1333 IRAS 4
\citep{atta09}. Here, Figure 8 shows the
result of fitting the observed spectral profile with the modified
infall model. Assuming $T_{\rm b,out}=50$ K and $V_{\rm out}=2$ km
s$^{-1}$, the best fitting gives $\tau_0=4.1\pm0.4$, $T_k=32\pm5$ K,
$\sigma=0.64\pm0.04$ km s$^{-1}$, $V_{\rm in}=0.86\pm0.06$ km s$^{-1}$,
$V_{\rm sys}=-38.6\pm0.1$ km s$^{-1}$ for the infall component (the
dashed profile in Figure 8) and $\tau_{\rm 0,out}=0.2\pm0.02$ and
$\sigma_{\rm out}=1.7\pm0.1$ km s$^{-1}$ for the outflow (dotted
profile). The spectral dip cannot be produced with a simple
superimposition of the double-Gaussian outflow profile (inset (a) in
Figure 8) to the infall profile (inset (b) in Figure 8). The spectral
dip seen in the resulting line profile is due to the absorption of the
cold infall gas located in front of the outflow. After solving the
radiative transfer equations, the final resulting profile of the model
fits the data reasonably well.

The result for the outflow component is not unique. Other combinations
of the three parameters $T_{\rm b,out}$, $V_{\rm out}$ and $\tau_{\rm
0,out}$ are also possible to produce a good fit to the observed line
profile. In addition to the second kinematic component due to outflow,
any other kinematic components within the CARMA beam ($\sim$6$\arcsec$)
characterized with two Gaussians separated in velocity, such as two
sub-cores orbiting around the mass center, can be incorporated into the
model in a way similar to that of the outflow, i.e., the dynamics in
the central 6$\arcsec$ may be equally important in the production of
the observed spectral profile. Furthermore, we also note that the
observed asymmetric line profile can be produced by an outflow alone if
an unequal optical depth is used for red- and blue-shifted components
in the outflow. Although the modeling is suggestive, we need improved
observations with high-angular resolution to differentiate between the
dynamical processes.

The $V_{\rm sys}$ = --38.6 km s$^{-1}$ derived from the model fitting
differs from the mean value (--38.9 km s$^{-1}$) of the peak velocities
from the single-peak optically thin lines such as C$^{18}$O (1-0)
observed with single-dish \citep{wu07}. Such a mean peak velocity
derived from observations of optically thin molecular lines is closer
to the systematic velocity of the cores. The small discrepancy between
the values of $V_{\rm sys}$ predicted by our model and derived from
observations might be due to the assumption of equal optical depth in
the blue- and red-shifted outflows used in the model. Furthermore, the
absorption dip in the observed HCO$^+$(1-0) spectrum is centered at
$\sim$--38.0 km s$^{-1}$, red-shifted relative to the systematic
velocity (--38.9 km s$^{-1}$) by 0.9 km s$^{-1}$, which is in good
agreement with the value of $V_{\rm in}$ derived from the model
fitting.

With the caveats discussed above, we can calculate the kinematic infall
rate $dM/dt$ = $4\pi r^2m_{\rm H}\mu n_{\rm nr}V_{\rm in}$ assuming the
spherical geometry. Given a radius $r$ = 6$\times$10$^3$ AU for the
main core A, a critical density of HCO$^+$(1-0) $n_{\rm nr}$ =
1.8$\times$10$^5$ cm$^{-3}$ \citep{evan99} and a mean molecular weight
$\mu$ = 2.35, an infall rate $dM/dt$ = 9.7$\times$10$^{-5}$ M$_\sun$
yr$^{-1}$ is inferred. This infall rate appears to be consistent with
the global gravitational rate $a^3/G$ = 6.8$\times$10$^{-5}$ M$_\sun$
yr$^{-1}$ \citep{shu77} assuming that the isothermal sound speed $a$
equals to the resulting velocity dispersion $\sigma$ (0.6 km s$^{-1}$).
Indeed, the sound speed of ideal H$_2$ gas at $T$ = 40 K is
$\sqrt{1.4kT/m_{\rm H_2}}$ = 0.5 km s$^{-1}$, consistent with the value
of the fitted $\sigma$.

Alternatively, rotation along with other dynamical process and the
depletion of HCO$^+$ in the range of velocity close to the systematical
velocity might be also responsible to the observed asymmetric
double-peak line profile \citep{pavl08}. Especially, the velocity
gradient as indicated in the P-V diagram of the optically thin
C$^{18}$O(1-0) line \citep[see][]{wu07} suggests that a rotation may
indeed play an important role in formation of the observed line
profile. In addition, although core A has been resolved into two
continuum cores in the SMA observations, this multiplicity does not
account for the observed double-peak line profile since further
molecular-line observations with higher angular resolution show that
the double-peak line profile exists in both of the cores \citep{zhu09}.

\section{Discussion}

\subsection{The Dynamical Age and Mass of the Outflows}

The dynamical age of the outflow can be inferred from the spatial
extension and the mean velocity of the outflow. In the case of W3-SE,
the maximum relative velocities ($>$5 $\sigma$) of the blue- and
red-shifted outflow lobes from the core A are $\sim$6 km s$^{-1}$ (see
the P-V diagram in Figure 7(b)) with respect to the systematic velocity
of --$38.9$ km s$^{-1}$ \citep{wu07}. Assuming a size $\sim$
20$\arcsec$ ($\sim$4$\times$10$^4$ AU) for each lobe, the dynamical age
of the outflow is $\sim$3$\times$10$^4$ yr. Since the major axis of the
outflow is poorly determined and its direction might be close to the
line of sight, the estimated dynamical age corresponds to a low limit.
Given the non-detection of radio continuum, both the typical outflow
velocity and the dynamical age indicate that W3-SE probably hosts
intermediate- and/or low-mass stars.

We calculated the mass associated with the main outflow in W3-SE using
the formula for column density in the case of a linear, rigid rotor
molecule on the assumption of LTE \citep{scov86}. For HCO$^+$(1-0), the
total column density at all energy levels can be estimated following
the equation:
\begin{eqnarray}
N({\rm HCO^+}) &=& \frac{3k}{8\pi^3B\mu_d^2}
(T_{\rm ex}+\frac{hB}{3k})\times \nonumber \\
&&(1-e^{-\frac{hv}{kT_{ex}}})^{-1}\int{\tau_{v}}dv
\end{eqnarray}
where we adopted the rotational constant $B$ = 44.594 GHz and the
permanent electric dipole moment $\mu_{\rm d}$ = 3.30 debye
\citep{wood75}. Assuming optically thin gas and excitation temperature
of T$_{\rm ex}$ = 40 K, the optical depth $\tau$ is proportional to the
brightness temperature $T_B$ \citep{wils09}:
\[
\tau = \frac{kT_B}{h\nu}(\frac{1}{e^{h\nu/kT_{ex}}-1}-
\frac{1}{e^{h\nu/kT_{bg}}-1})^{-1}
\]

The outflow mass can be calculated from Equation (3) in the case of optically
thin emission, which valid for the outflow gas in the high-velocity
wings,
\begin{eqnarray}
M({\rm H_2}) &=& X_{\rm HCO^+}^{-1}m({\rm H_2})D^2 \times \nonumber \\
&&\int{N({\rm HCO^+})}d\Omega
\end{eqnarray}
where we adopted a fractional abundance X$_{\rm HCO^+}$ =
7.6$\times$10$^{-9}$ for W3-SE \citep{zinc09} and a distance $D$ = 2
kpc. Then, we determined the velocity-integrated brightness temperature
averaged over the emission region $\Omega$ $\sim$
50$\arcsec$$\times$50$\arcsec$, $\int{<T_B>_\Omega}{\rm d}v$ = 1.2 K km
s$^{-1}$ between --46.7 and --41.8 km s$^{-1}$, and
$\int{<T_B>_\Omega}{\rm d}v$ = 1.4 K km s$^{-1}$ between --36.5 and
--31.6 km s$^{-1}$. The corresponding masses are 2.5 and 2.8 M$_\sun$
for the blue- and red-shifted outflow lobes, respectively.
Consequently, with the maximum relative velocities of the outflow lobes
($\sim$6 km s$^{-1}$) and the dynamic age ($\sim$3$\times$10$^4$ yr),
the momentum rates for both the blue and red lobes are
$\sim$5$\times$10$^{-4}$ M$_\sun$ km s$^{-1}$ yr$^{-1}$, corresponding
to mechanical luminosities of 0.2 L$_\sun$. The inferred outflow
momentum rates and mechanical luminosities appear to be consistent with
B-type stars in W3-SE \citep{arce06}.

\subsection{A Group of Young Stars/Protostars in W3-SE}

Observations at both millimeter and submm wavelengths, together
with those at mid-IR suggest that the main molecular core A hosts a
group of young stars and/or protostars. The millimeter and
submm observations suggest that the cold dust is distributed in
an elongated region from SE to NW (Figure 2) with two dust condensations
separated by 4$\arcsec$. The non-detection of radio continuum at
$\lambda$~6 cm \citep{tief98} indicates that no OB stars have been
found in W3-SE yet although we cannot rule out the possibility that the
two sub-cores in the molecular core A (SMA-1 and -2) will continue
to accrete from the molecular gas reservoir of $\sim$65 M$_\sun$ and
form a massive star. The four IRAC images show a small group of
discrete sources in the dust core. One (IRAC 56) of them has been
identified by \cite{ruch07} as a point source, suggesting an
association with a YSO earlier than Class I. This source appears to be
one of the youngest object in the entire W3 cluster as indicated in the
color-color diagram of \cite{ruch07}. The weak point-like source IRAC
56a might be associated with a protostar embedded in the dust core
SMA-1.

The positions of the emission peaks in the millimeter/submillimeter
continuum and lines appear to be displaced significantly from the
brightest mid-IR sources in this region. The majority of the mid-IR
emission arises from the region located east of the molecular core (see
Figure 2). The displacement likely indicates that the hot and cold dust
components which are inferred from fitting the SED are also spatially
separated in the W3-SE region. The spatial separation between the hot
and cold dust might be explained by sequential star formation in W3-SE.
Namely, stars formed first in the region east to the main molecular
core A in W3-SE, where the dust/molecular components might have been
substantially dissociated or destroyed by the stars. The activity of
forming new stars is now taking place in A. The activity of star
formation appears to propagate from E to W in W3-SE. Alternatively, the
displacement of the emission peak position at mid-IR from that of
millimeter/submillimeter could be due to possible strong mid-IR
extinction in the main molecular core A. The scenario of sequential
activities in star formation needs to be verified by further improved
mid-IR observations with corrections for dust extinction.

\section{Summary and Conclusions}

We observed W3-SE with CARMA and SMA at $\lambda$~3.4 and 1.1 mm. Using
the continuum flux densities determined with CARMA, SMA and JCMT
together with measurements at mid-IR wavelengths with {\it Spitzer} and
MSX, we constructed the SED for the W3-SE core. The SED can be fitted
with two thermal dust components, suggesting that the majority of the
gas/dust (65$\pm$10 M$_\sun$) is at a relatively low temperature of
41$\pm$6 K corresponding to radiation peaked at a far-IR wavelength.
There is an additional hot component ($\sim$400 K) with little mass
accounting for the secondary emission bump at mid-IR in the SED,
suggesting that a small fraction of dust has been heated by newly-born
stars in the immediate surroundings.

The CARMA observations of the HCO$^+$(1-0) line confirmed the presence
of blue-shifted double-peak line profile towards the W3-SE core
discovered by \cite{wu07} with the IRAM-30m. We made an attempt to fit
the line profile with the infall model of \cite{myer96} by
incorporating outflow components. The model fitting suggests that in
addition to infalling gas, a significant contribution from outflows is
necessary to produce the observed line profiles. The kinematics of the
molecular gas in W3-SE observed with CARMA in the central 40$\arcsec$
region are consistent with a bipolar outflow with a dynamic age of
$\sim$3$\times$10$^4$ yr. Finally, based on the observations with
CARMA, SMA, and JCMT along with the previous published results, we
conclude that W3-SE hosts a group of young stars and protostars which
are likely located in the double dust cores observed with SMA.

\acknowledgments

We are grateful for the staff of the CARMA, SMA and JCMT for their
assistance in the observations. L. Z. thanks to the organizers and
tutors of the CARMA 2008 summer school, and J. Carpenter for his help
in data reduction. We are grateful for G. Ruch for providing the IRAC
mosaic images and the point-source table, and the useful discussions
with us on the infrared part of this work. L. Z. is supported by the
SAO predoctoral program. L. Z. thanks the NRAO for providing him NSF
travel fund to attend the mm/submm conference in Taipei. Y. W. and
L. Z. are grateful for the support of the Grant 10873019 of NSFC.

\clearpage
\begin{deluxetable*}{lccccccccc}
\tablecaption{Calibrators for the CARMA and SMA Observations}
\tablewidth{0pt} \tablehead{\colhead{Date}&\colhead{$\lambda$}
&\multicolumn{2}{c}{Bandpass}&&\multicolumn{2}{c}{Phase}
&&\multicolumn{2}{c}{Flux Density} \\
\cline{3-4} \cline{6-7} \cline{9-10}
&(mm)&\colhead{Source}&\colhead{Flux(Jy)}&&\colhead{Source}&
\colhead{Flux(Jy)}&&\colhead{Source}&\colhead{Flux(Jy)}} \startdata
2008-06-24    &3.4 &3C84    &6.1  &&0102+584 &2.1 &&Uranus &7.2  \\
2008-06-27    &3.4 &3C454.3 &27.8 &&0102+584 &2.5 &&MWC349 &1.3  \\
2008-10-27    &1.1 &3C454.3 &13.2 &&0244+624 &0.6 &&Uranus &45.6
\enddata
\end{deluxetable*}

\begin{deluxetable*}{ccccccc}
\tablecaption{Measurements of the Continuum at mm/submm}
\tablewidth{0pt} \tablehead{
$\lambda$     &Telescope &\multicolumn{4}{c}{Gaussian Fitting Results} &S$_{\nu}$   \\
\cline{3-6} (mm)          & &Peak(\arcsec) &I$_{\rm peak}$(Jy
beam$^{-1}$) &Sizes(\arcsec) &P.A.(\arcdeg) &(Jy)
 } \startdata
3.4           &CARMA   &--5.5,--0.9 &0.024$\pm$0.002 &7.9$\times$5.9 &-51 &0.04$\pm$0.01 \\
1.1           &SMA     &---         &---             &---            &--- &1.4$\pm$0.4   \\
              &(SMA-1) &--6.2, 0.2  &0.21$\pm$0.01   &4.4$\times$3.0 &21  &0.72$\pm$0.01 \\
              &(SMA-2) &--4.2,--2.8 &0.17$\pm$0.01   &3.5$\times$1.5 &76  &0.35$\pm$0.01 \\
0.85          &JCMT    &--8.2, 1.2  &3.3$\pm$0.1     &25$\times$18   &-48 &8$\pm$1       \\
0.45          &JCMT    &--8.8, 0.5  &17.1$\pm$0.7    &18$\times$11   &-41 &75$\pm$8
\enddata
\tablecomments{Columns 1 and 2 give the wavelengths and telescopes
which are used in the observations. Columns 3, 4, 5 and 6 are the
results from Gaussian fitting to the sources, corresponding to peak
positions, peak intensities, core sizes and position angles. Column 5
gives the total flux densities integrating the emission $>$5 $\sigma$
over the continuum core region of 15$\arcsec$$\times$15$\arcsec$
(except for SMA-1 and -2). The uncertainties are attributed mainly to
the missing short spacing data for the interferometer observations and
the uncertainty in determination of the boundary for the continuum core
in the JCMT images. For SMA-1 and SMA-2, the total flux densities are
determined from multiple Gaussian fitting to the 1.1-mm dust emission.}
\end{deluxetable*}

\begin{deluxetable*}{ccc}
\tablecaption{The Continuum Flux Densities at mid-IR}
\tablewidth{0pt}
\tablehead{
$\lambda$        &Instrument &S$_{\nu}$ \\
(\micron)        &          & (Jy)   }
\startdata
24               &MIPS  & 8.8$\pm$1.3     \\
21.3             &MSX   & 9.2$\pm$3.7     \\
14.7             &MSX   & 0.83$\pm$0.33      \\
12.1             &MSX   & 0.94$\pm$0.38      \\
8.3              &MSX   & 0.95$\pm$0.38      \\
8.0              &IRAC  & 0.82$\pm$0.16      \\
5.8              &IRAC  & 0.48$\pm$0.10      \\
4.5              &IRAC  & 0.18$\pm$0.04    \\
3.6              &IRAC  & 0.054$\pm$0.011
\enddata
\tablecomments{The observing wavelengths, the relevant instrument and
the continuum flux densities are given in columns 1, 2 and 3,
respectively.}
\end{deluxetable*}

\begin{figure}[h]
  \centering
  \includegraphics[width=6cm, angle=-90, origin=c]{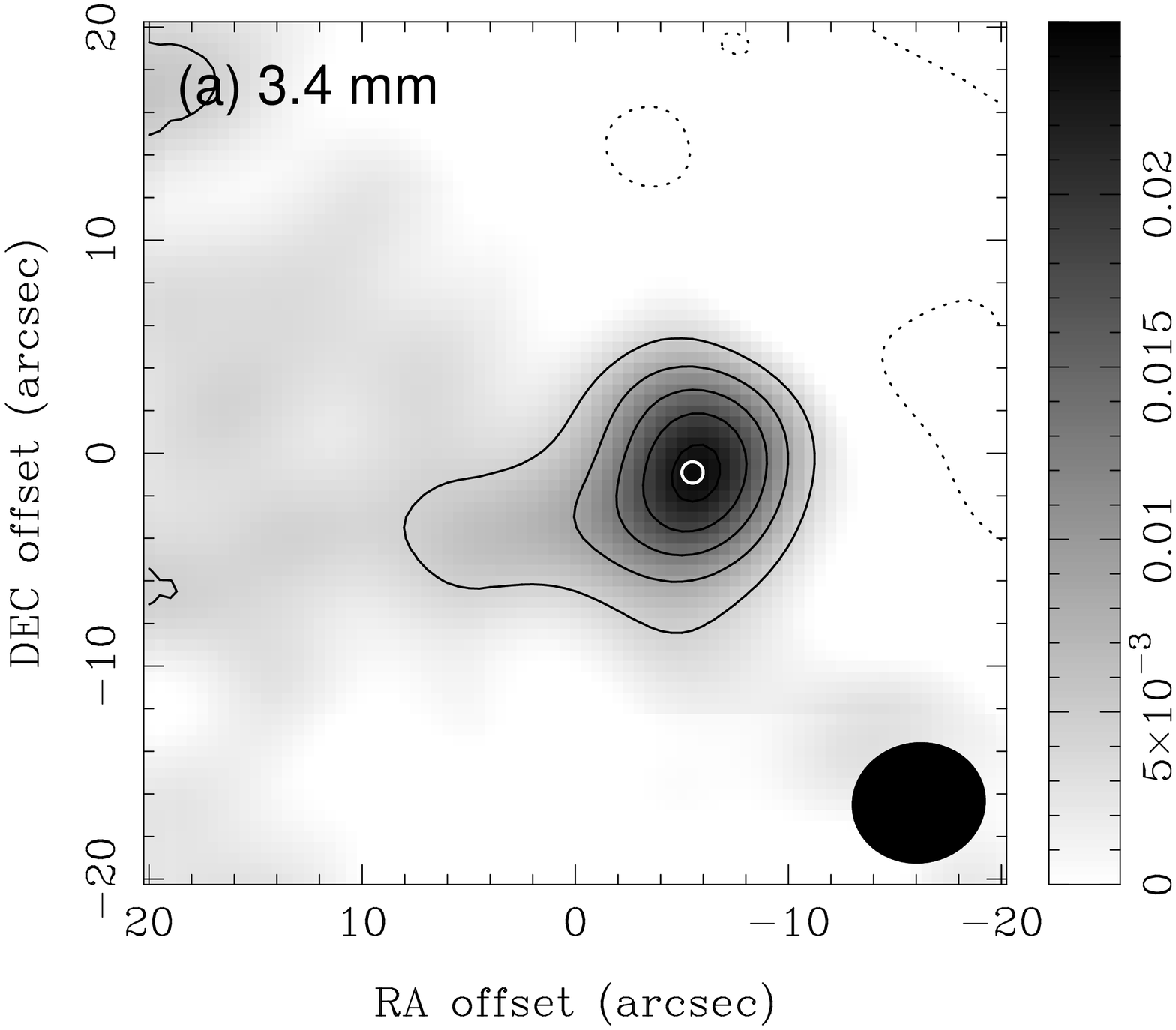}
  \hspace*{1cm}
  \includegraphics[width=6cm, angle=-90, origin=c]{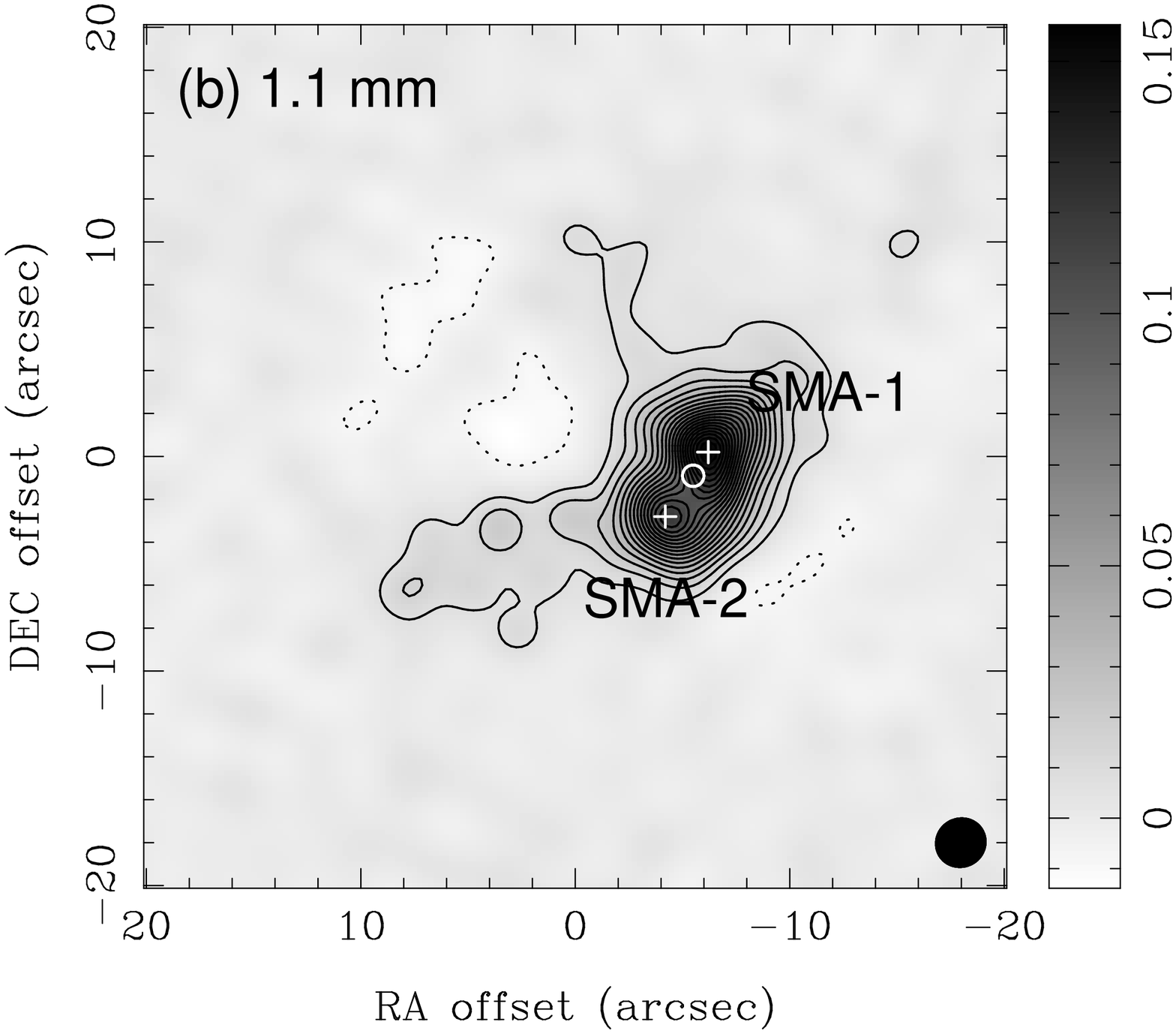}
  \\
  \includegraphics[width=6cm, angle=-90, origin=c]{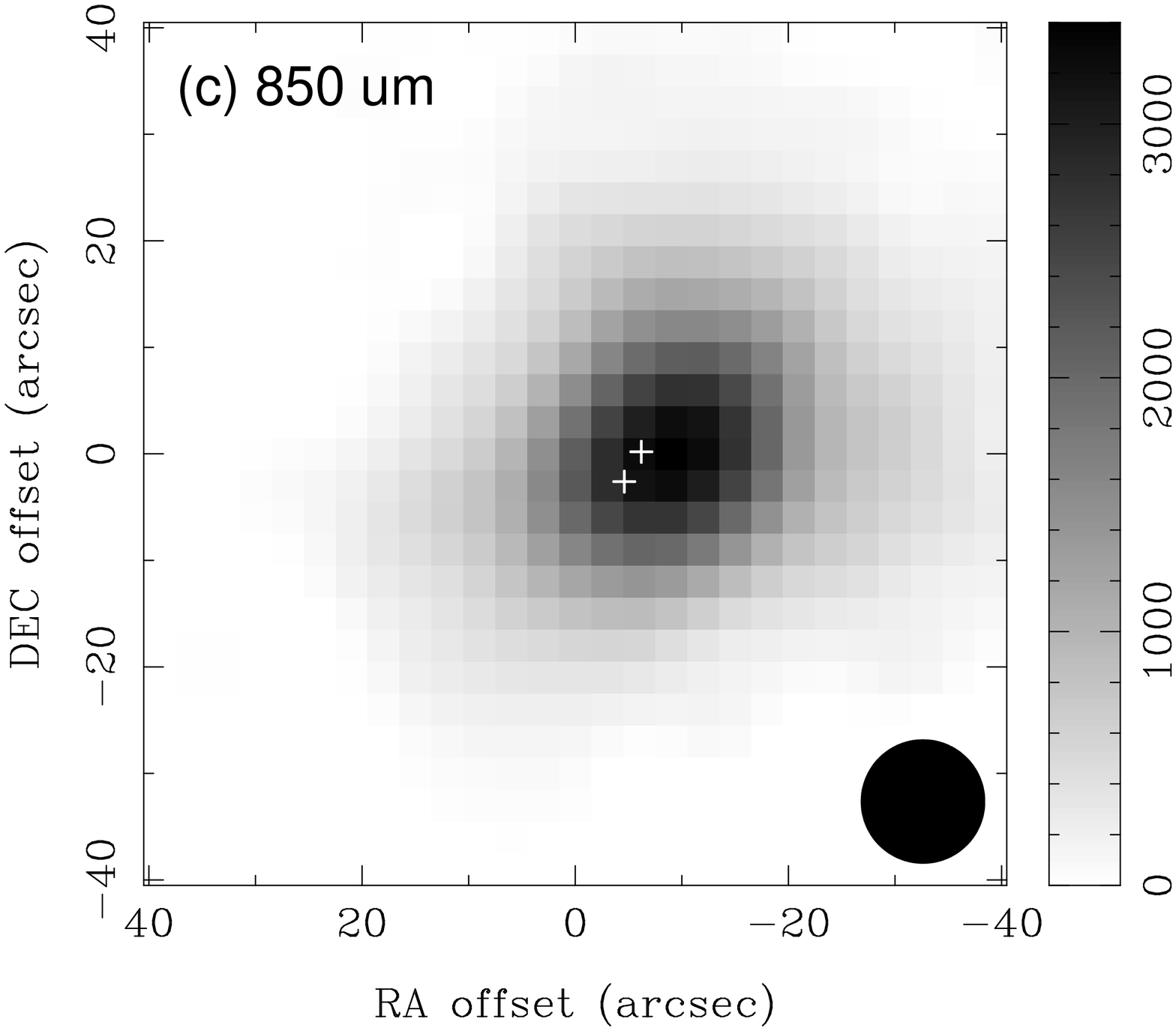}
  \hspace*{1cm}
  \includegraphics[width=6cm, angle=-90, origin=c]{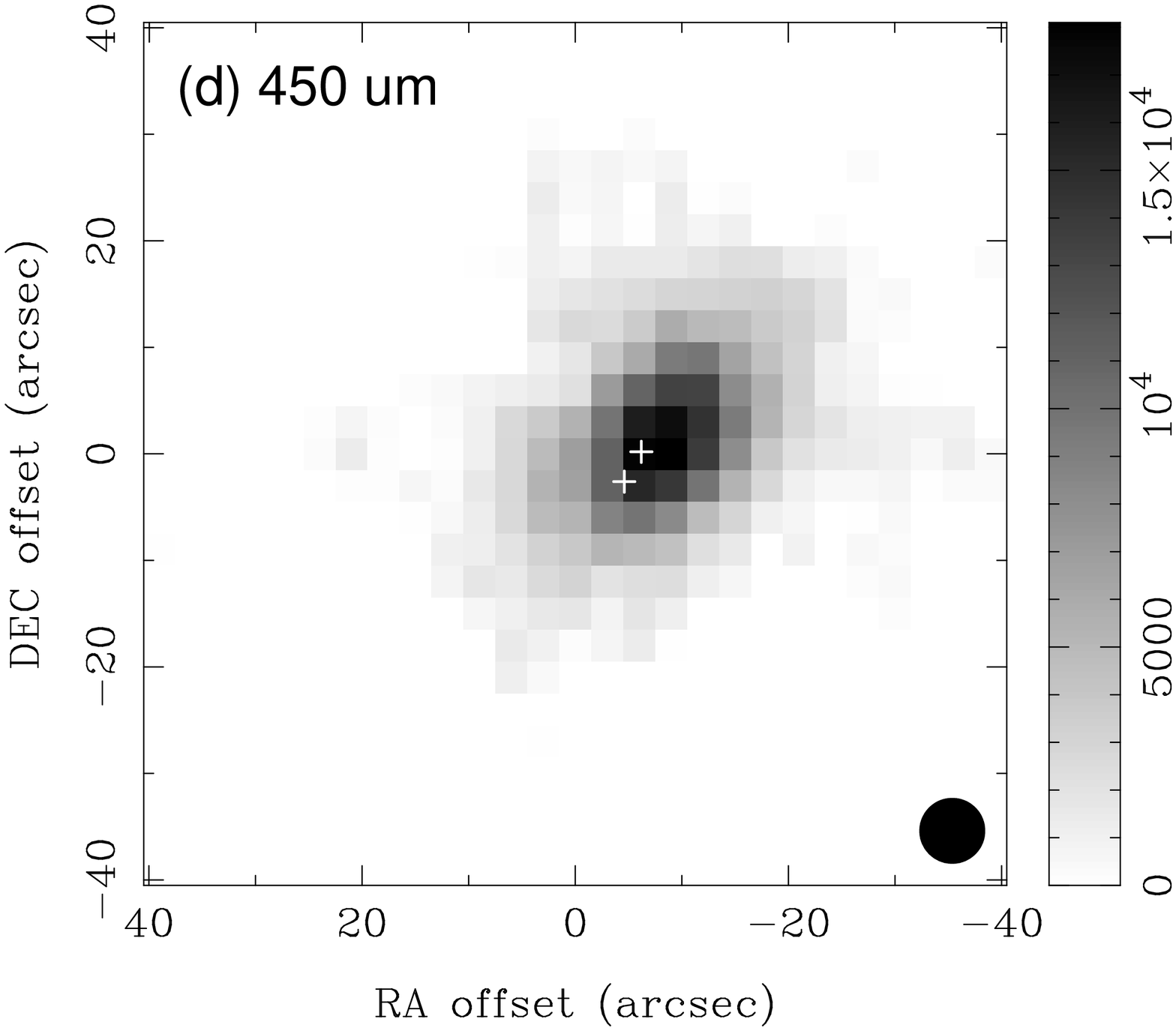}
  \caption{The dust emission of W3-SE. (a) The image of the continuum
  emission at $\lambda$~3.4 mm made from the CARMA observations by
  combining two 500-MHz bands. The contours are 3$\sigma$$\times$
  (-1, 1, 2, 3, ...) and 1 $\sigma$ = 0.0015 Jy beam$^{-1}$. The FWHM
  beam size given at bottom-right is 6.2$\arcsec$$\times$5.6$\arcsec$,
  P.A.=--81$\arcdeg$. (b) The image of continuum emission at 1.1-mm
  based on the SMA observations in the compact array with two 2-GHz
  side bands. The contours are 5$\sigma$$\times$ (-1, 1, 2, 3, ...)
  and 1 $\sigma$ = 0.0015 Jy beam$^{-1}$. The FWHM beam size is
  2.9$\arcsec$$\times$2.7$\arcsec$, P.A. = --64$\arcdeg$. The open
  circle in both (a) and (b) marks the peak position of the 3.4-mm
  emission. The images of continuum emission at 0.85 mm (c) and
  0.45 mm (d) were made with the JCMT observations with the FWHM beam
  sizes of 14.5$\arcsec$ and 7.5$\arcsec$, respectively. The crosses
  in (b), (c) and (d) mark the peak positions of the double sources
  (SMA-1 and -2). The units of the wedge in (a) and (b) are Jy
  beam$^{-1}$; mJy beam$^{-1}$ for (c) and (d).}
\end{figure}

\begin{figure}[h]
  \centering
  \includegraphics[width=5cm, angle=-90, origin=c]{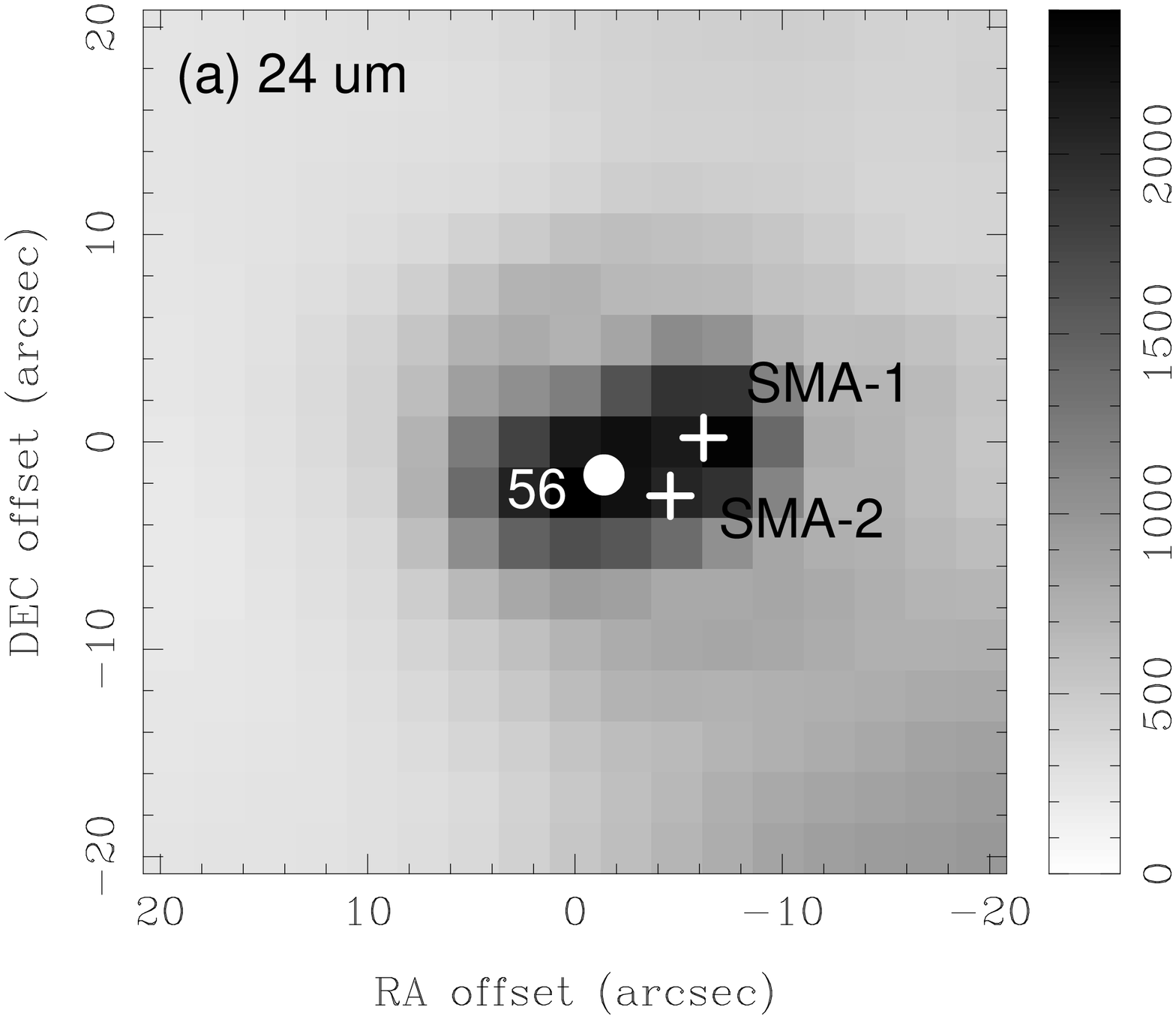}
  \hspace*{1cm}
  \includegraphics[width=5cm, angle=-90, origin=c]{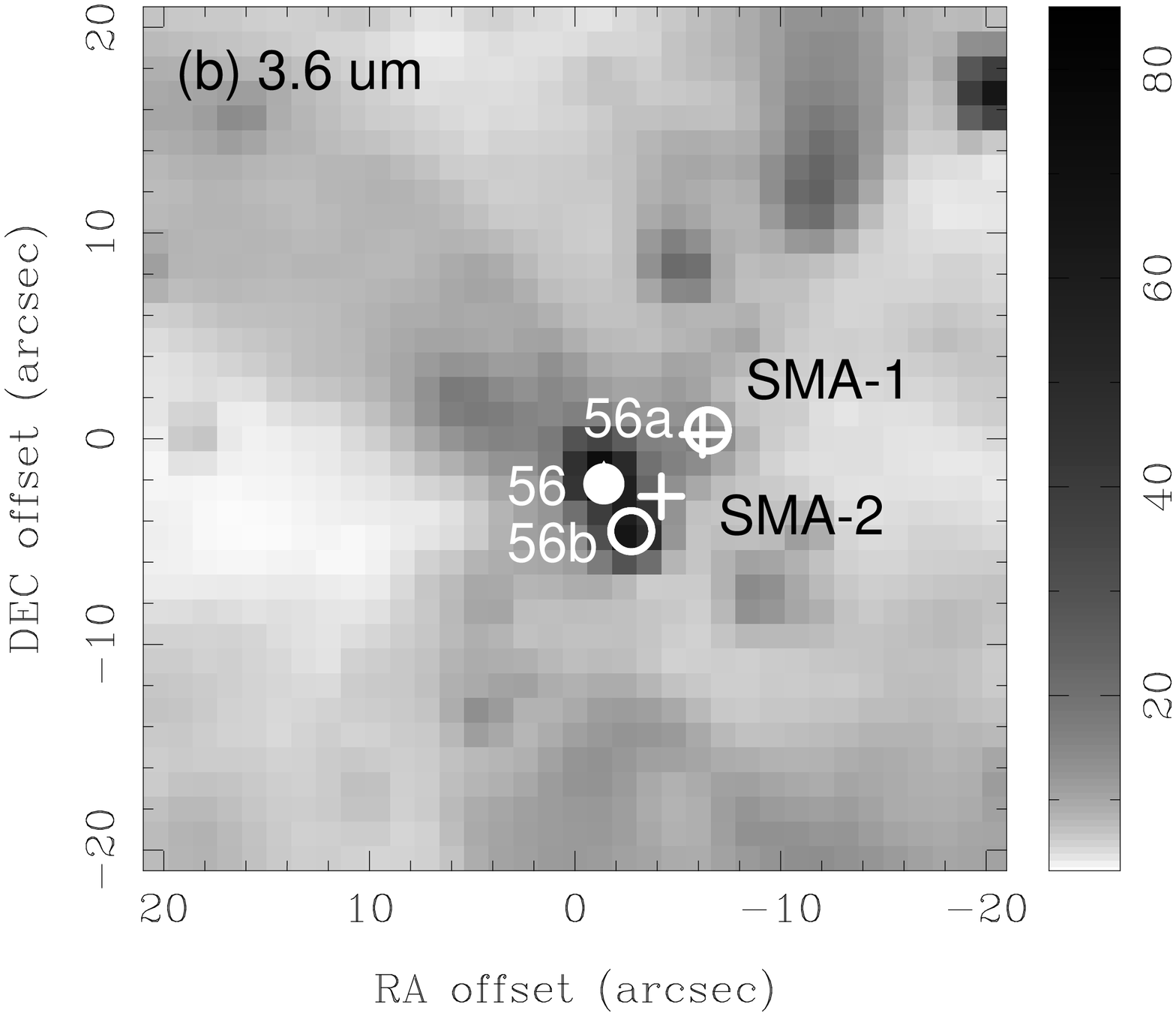}
  \caption{{\it Spitzer} images at (a) MIPS 24 $\micron$ and
  (b) IRAC 3.6 $\micron$. In the two images the wedge scales
  the continuum emission in units of 10$^6$Jy sr$^{-1}$,
  which is equivalent to 2.35$\times$10$^{-5}$ Jy arcsec$^{-2}$.
  The crosses in the two images indicate the peak positions of
  the double sources. The filled circle in (a) and (b) marks
  position of the mid-IR point source (IRAC 56) which
  has been identified by \cite{ruch07}. The open circles
  in (b) indicate two additional sources (IRAC 56a and 56b) in
  the central core region of W3-SE.
  }
\end{figure}

\begin{figure}[h]
  \centering
  \includegraphics[width=6cm, angle=0]{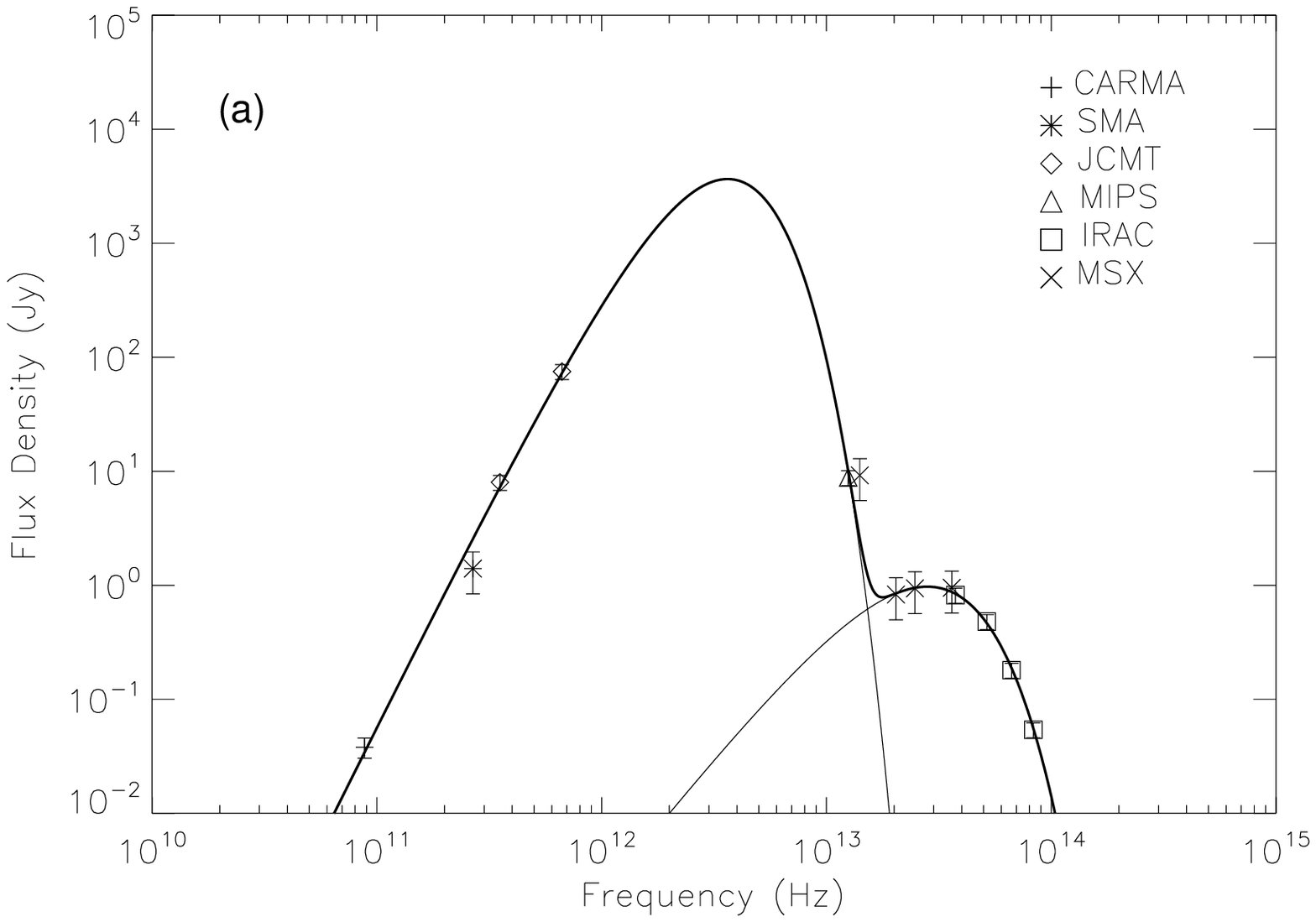}
  \includegraphics[width=6cm, angle=0]{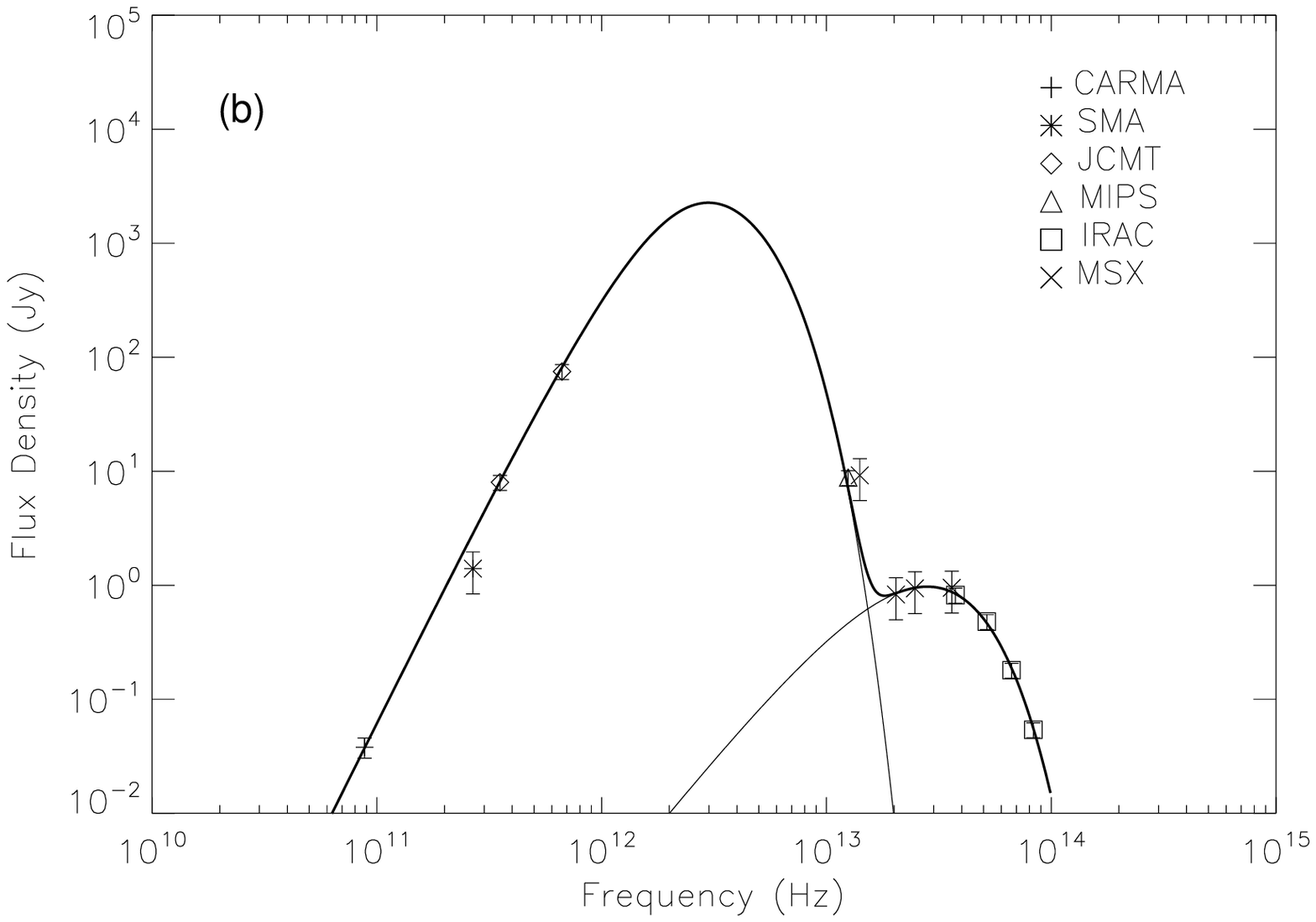}
  \caption{The SED of the continuum emission from the W3-SE core
  determined from the observations with the CARMA, SMA, JCMT, MSX,
  MIPS and IRAC. The legend (top-right) shows the symbols
  corresponding to each of the telescopes. The thick solid curve
  outlines the resulting SED fitting from two thermal dust components
  (indicated by the thin curves) determined from our model fitting.
  (a) The best (LSQ) fit to the flux density measurements of the SED
  assuming the optically thin dust emission from both cold and hot
  components. (b) A good fit to the SED considering the effect of
  self-absorption of the cold component with $\beta=2$ and assuming
  that the cold component does not overlap the hot dust emission region.
  The turnover frequency $\nu_{\tau=1}$ of
  the cold component is shifted by $\Delta\nu$ = 80 GHz to higher
  frequency as compared to that derived from the optically-thin approach.}
\end{figure}

\begin{figure}[h]
  \centering
  \vspace*{2cm}
  \hspace*{-5cm}
  \includegraphics[width=12cm, angle=-90]{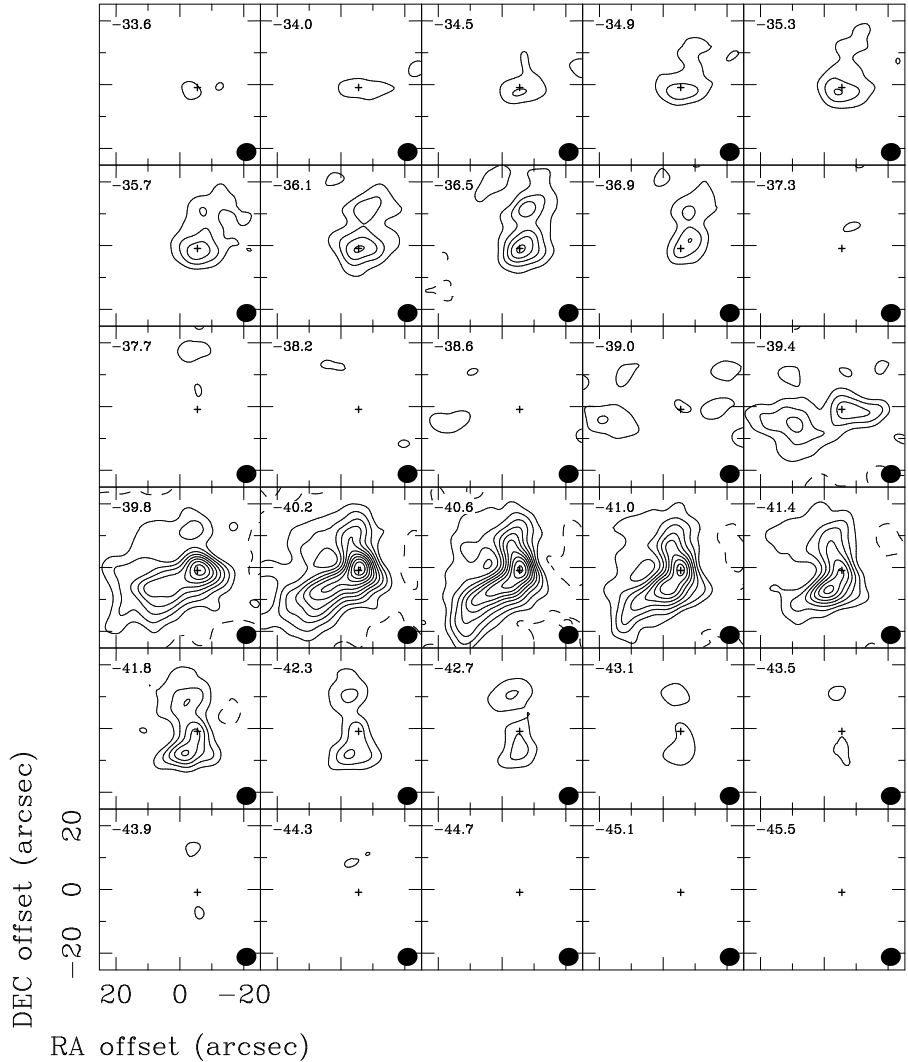}
  \caption{Channel maps of the HCO$^+$(1-0) line observed with
  the CARMA. The channel separation is 0.42 km s$^{-1}$. The
  contours are 4$\sigma$$\times$(-1, 1, 2, 3, ...) and 1 $\sigma$
  = 0.1 Jy beam$^{-1}$. The LSR velocity is labeled at top-right
  on each channel map. The FWHM beam (bottom-right) is
  6.3$\arcsec$$\times$5.7$\arcsec$, P.A.=--79$\arcdeg$.}
\end{figure}

\begin{figure}[h]
  \centering
  \vspace*{2cm}
  \hspace*{-5cm}
  \includegraphics[width=9.0cm, angle=-90, origin=c]{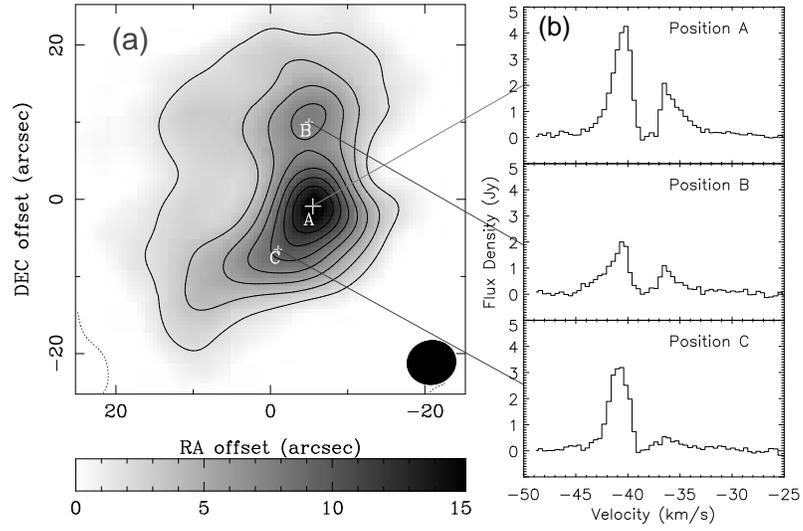}
  \caption{(a) The integrated intensity map of HCO$^+$(1-0)
  from the CARMA observations. The contours are
  3$\sigma$$\times$(-1, 1, 2, 3, ...) and 1 $\sigma$ = 0.6
  Jy beam$^{-1}$ km s$^{-1}$. The FWHM beam size is
  6.3$\arcsec$$\times$5.7$\arcsec$, P.A. = --81$\arcdeg$,
  shown at bottom-right. The cross marks the
  $\lambda$~3.4-mm continuum peak. (b) The spectra towards
  positions A, B and C marked in Figure 5(a).}
\end{figure}

\begin{figure}[h]
  \begin{minipage}[b]{0.5\linewidth}
  \centering
  \includegraphics[width=6.0cm, angle=-90, origin=c]{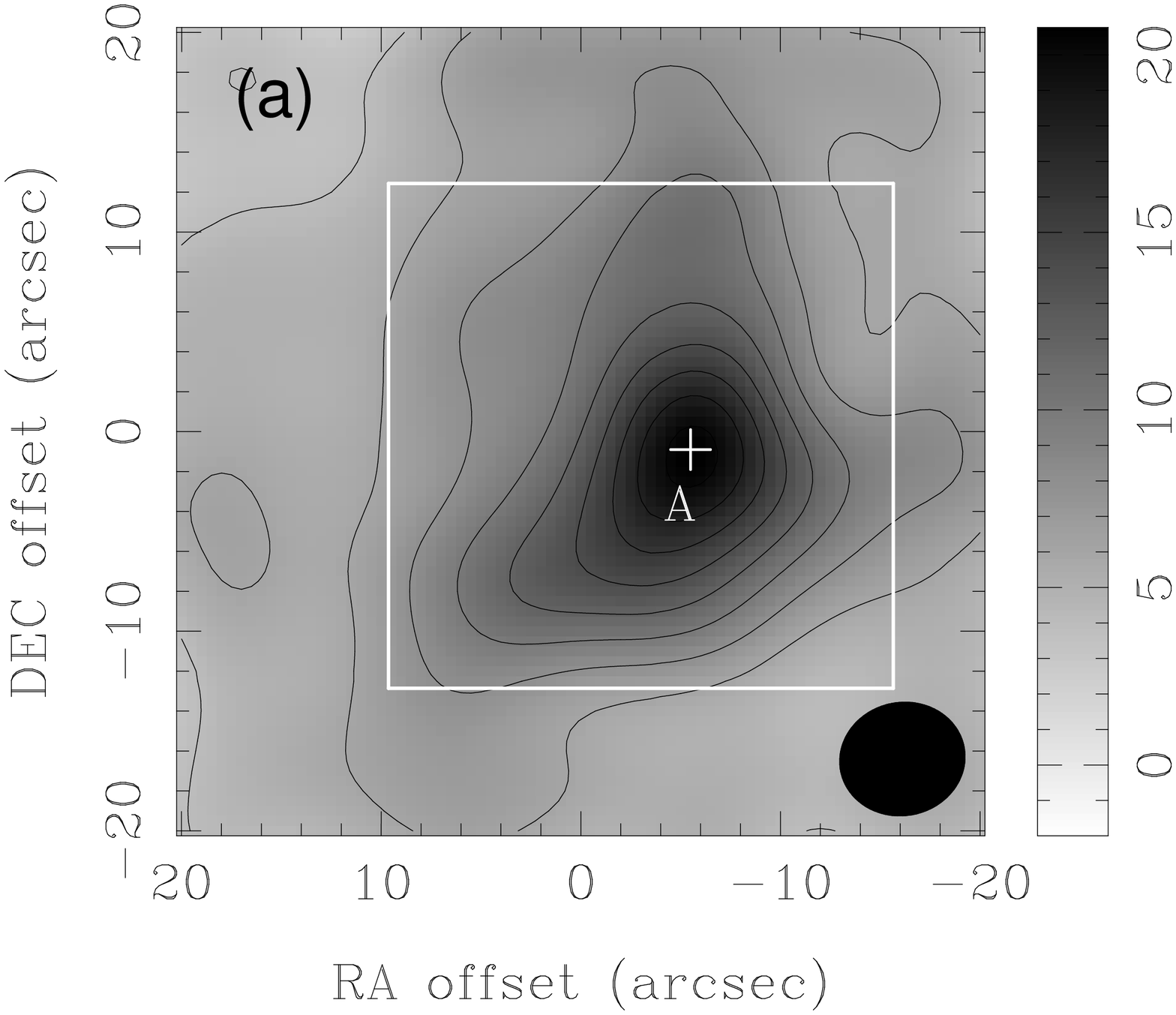}
  \end{minipage}
  \begin{minipage}[b]{0.5\linewidth}
  \centering
  \includegraphics[height=6.6cm, angle=0]{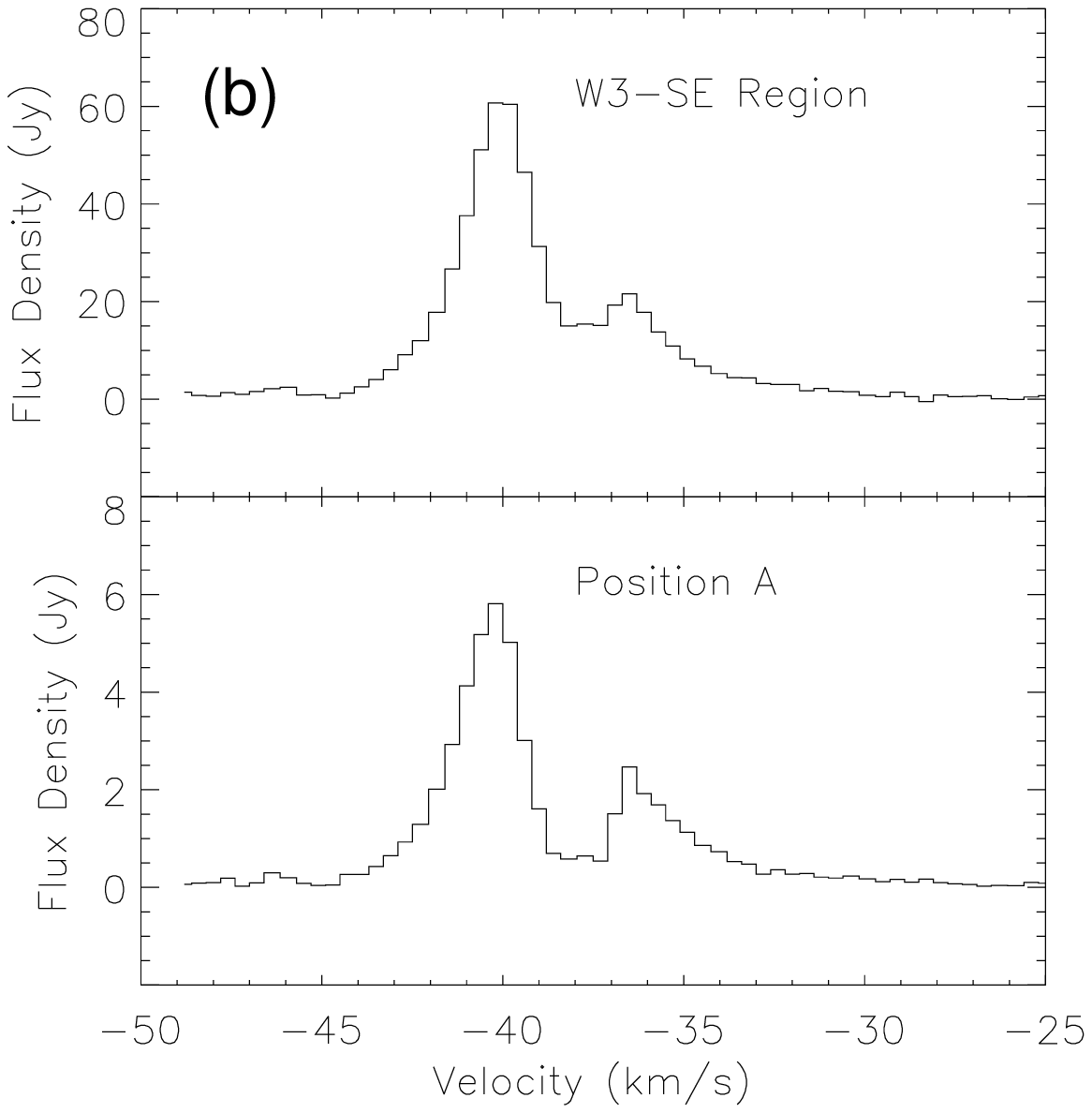}
  \end{minipage}
  \caption{(a) The integrated intensity image of the
  HCO$^+$(1-0) line emission constructed by combining the CARMA
  data with the IRAM 30-m data. The beam size of
  6.3$\arcsec$$\times$5.7$\arcsec$, P.A.=--81$\arcdeg$ is the same
  as the one of CARMA image (Figure 5(a)). The boundary effect has been
  discussed in the text. (b) The top spectrum is the integrated
  HCO$^+$(1-0) line emission within the 25$\arcsec\times25\arcsec$
  square as delineated. The spectrum toward the peak position A
  is shown at the bottom.}
\end{figure}

\begin{figure}[h]
  \centering
  \begin{minipage}[b]{0.45\linewidth}
  \centering
  \includegraphics[width=6cm, angle=-90, origin=c]{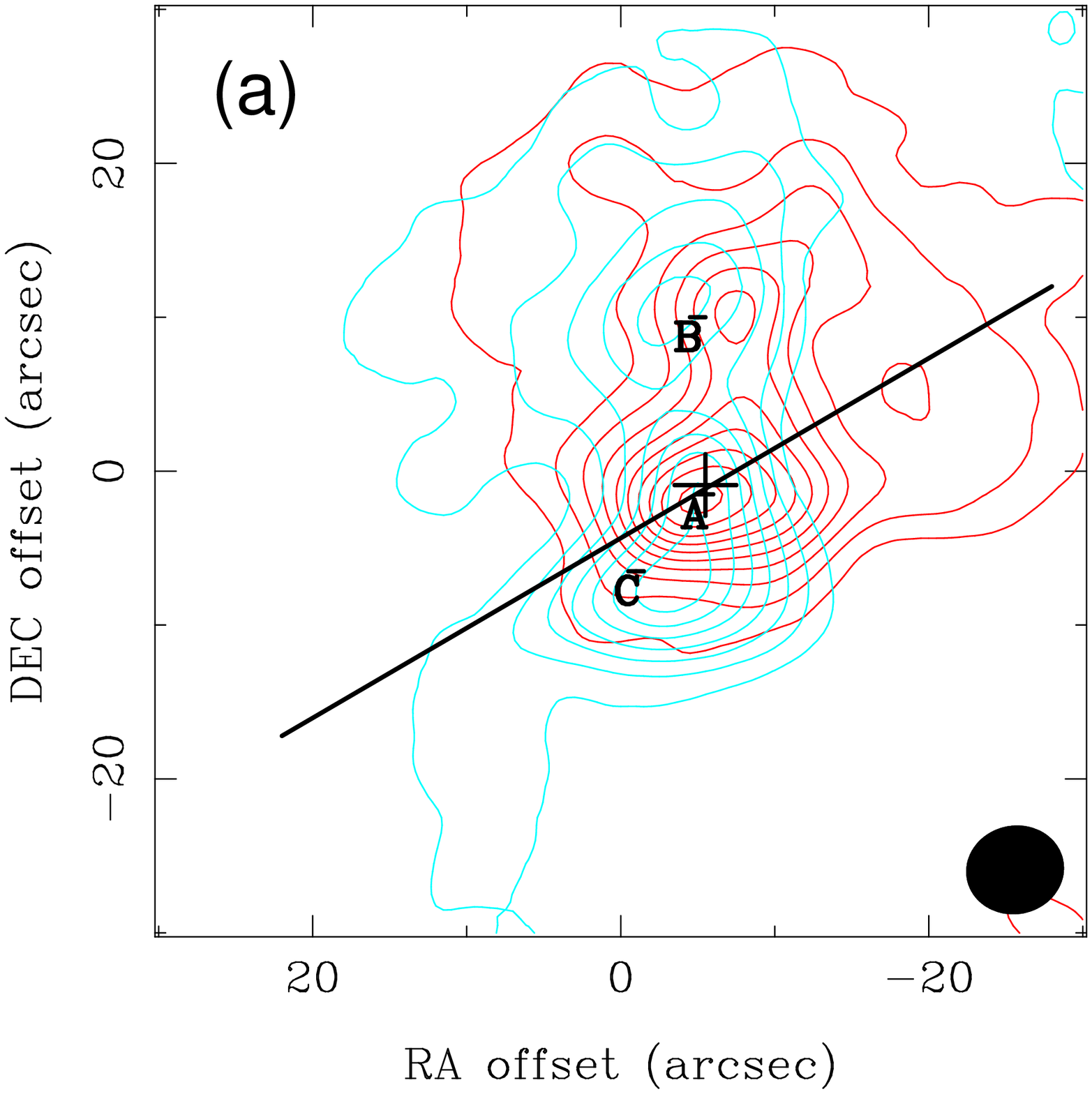}
  \end{minipage}
  \begin{minipage}[b]{0.45\linewidth}
  \centering
  \includegraphics[width=5cm, angle=-90, origin=c]{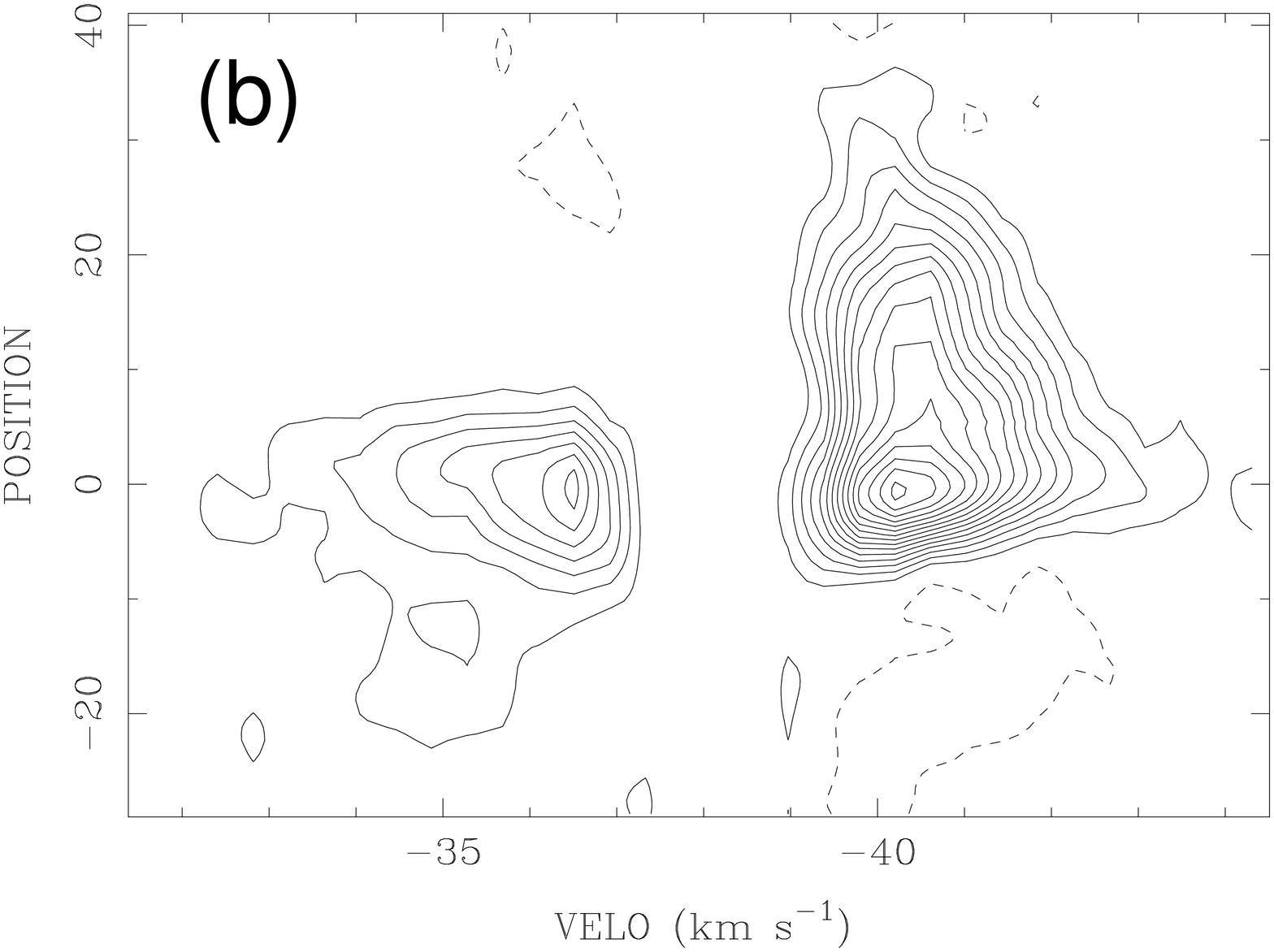}
  \vspace*{0cm}
  \end{minipage}
  \vspace*{-1cm}
  \caption{The integrated intensity maps for
  the blue and red wings based on CARMA data. The contours
  are 5$\sigma$$\times$(1, 2, 3, ...) and 1 $\sigma$ =
  0.12 Jy beam$^{-1}$ km s$^{-1}$. The FWHM beam size is
  6.3$\arcsec$$\times$5.7$\arcsec$, P.A.=--81$\arcdeg$.
  The straight line denotes the direction of the cut for
  the P-V diagram in (b). (b) The P-V diagram along
  the cut indicated in Figure 7(a). The vertical axis is the
  offset relative to the position A, in units of arcsec,
  positive to SE. The contours are
  5$\sigma$$\times$(1, 2, 3, ...) and 1 $\sigma$ =
  0.08 Jy beam$^{-1}$.}
\end{figure}

\begin{figure}[h]
  \centering
  \centering
  \includegraphics[height=8.0cm, angle=0]{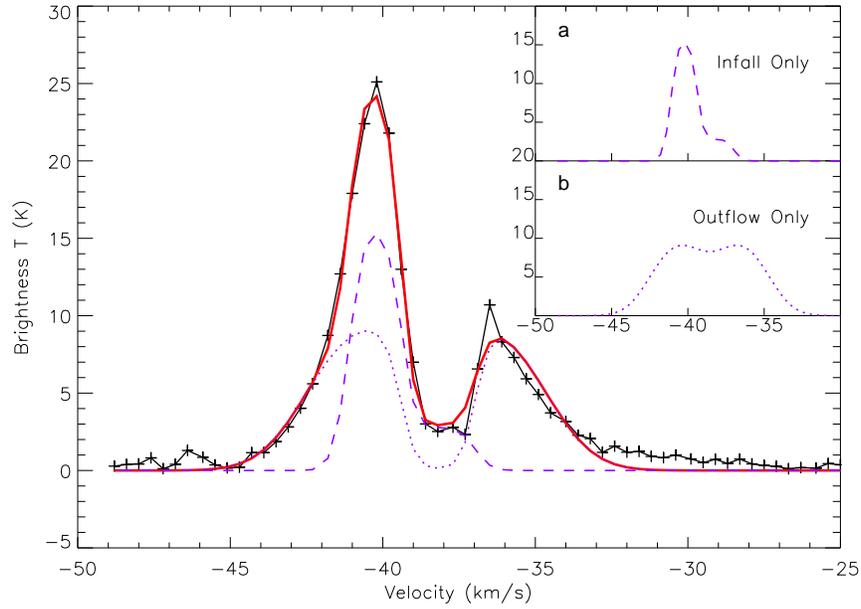}
  \caption{The observed line profile (crosses connected
  with a black curve) is fitted with a two-layer infall
  model including blue- and red-shifted outflow components.
  The solid curve (red) shows the resulting profile from
  the infall (dashed curve) and outflow (dotted curve)
  components. The insets at the top-right show
  the two spectra produced separately with infall alone (upper)
  and outflow alone (lower) using the same parameters derived
  from the overall spectra fitting.}
\end{figure}

\end{document}